\documentclass[aps,reprint,pra,amsmath,amssymb,superscriptaddress]{revtex4-1}
\usepackage{graphicx,wasysym}
\usepackage{dcolumn}
\usepackage{bm}
\usepackage{bbm}
\usepackage{natbib}
\usepackage{tabulary,tabularx}
\usepackage[export]{adjustbox}
\usepackage{color}
\usepackage[utf8]{inputenc}
\usepackage[T1]{fontenc}
\usepackage{float}
\usepackage{hyphenat}
\usepackage{hyperref}
\usepackage{xcolor}
\hypersetup{
  colorlinks   = true, 
  urlcolor     = blue, 
  linkcolor    = blue, 
  citecolor   = blue, 
}

\begin{document}

\title{Properties of the Superfluid in the Disordered Bose-Hubbard Model}

\author{Bruno R. de Abreu}
\email{bricardi@ifi.unicamp.br}
\affiliation{Instituto de F\'\i sica Gleb Wataghin,
University of Campinas - UNICAMP, 13083-859 Campinas - SP, Brazil}

\author{Ushnish Ray}
\email{uray@caltech.edu}
\affiliation{Division of Chemistry and Chemical Engineering, California Institute of Technology, Pasadena, CA 91125, USA}

\author{Silvio A. Vitiello}
\affiliation{Instituto de F\'\i sica Gleb Wataghin,
University of Campinas - UNICAMP, 13083-859 Campinas - SP, Brazil}
\affiliation{CENAPAD-SP,
University of Campinas - UNICAMP, 13083-889 Campinas - SP, Brazil}

\author{David M. Ceperley}
\affiliation{Department of Physics, University of Illinois at Urbana-Champaign, Champaign, IL 61801, USA}

\begin{abstract}
We investigate the properties of the superfluid phase in the three-dimensional disordered Bose-Hubbard model using Quantum Monte-Carlo simulations. The phase diagram is generated using Gaussian disorder on the on-site potential. Comparisons with box and speckle disorder show qualitative similarities leading to the re-entrant behavior of the superfluid. Quantitative differences that arise are controlled by the specific shape of the disorder. Statistics pertaining to disorder distributions are studied for a range of interaction strengths and system sizes, where strong finite-size effects are observed. Despite this, both the superfluid fraction and compressibility remain self-averaging throughout the superfluid phase. Close to the superfluid-Bose-glass phase boundary, finite-size effects dominate but still suggest that self-averaging holds. Our results are pertinent to experiments with ultracold atomic gases where a systematic disorder averaging procedure is typically not possible. 
\end{abstract}

\maketitle


The consequences of adding disorder to interacting particles present a very important problem for many-body physics. 
For Bosons, where interactions play a crucial role, the disordered Bose-Hubbard model (DBHM) allows one to study the interplay between correlation and disorder. Ever since the seminal work of Fisher et al. \cite{Fisher1989}, this model has received a lot of attention and  its properties have been explored using renormalization group (RG) approaches, numerical techniques, as well as experiment \cite{Ma1986,Fisher1989,Krauth1991,Scalettar1991,Pollet2009,White2009a,Micklitz2014,Pons2017,Paiva2015,Meldgin2016,Volchkov2018}. 

In order to understand the physics of DBHM, it is useful to first consider its clean counterpart, the Bose-Hubbard model (BHM), where Bosons are allowed to move on a lattice and interact. This system illustrates one of the most fundamental results of modern physics -- that of a quantum phase transition (QPT). By tuning the interaction strength, the system switches between two distinct phases corresponding to the two possible ground-states: the delocalized superfluid (SF) and the localized Mott-Insulating (MI) state \cite{Gersch1963,Gavish2005,Calzetta2006,Tilahun2011,Capogrosso-Sansone2007,Lacki2016}. The relatively recent innovation of synthetic materials, made possible with ultra-cold atomic gases, has added a vitally important tool to study many-body physics in experimental systems \cite{Lewenstein2007,Orzel2001,Bloch2008,White2009a,Zhang2012,Ohmori2017,Hofstetter2006,Esslinger2010}. Such systems have been used to study the SF-MI transition \cite{Greiner2002,Chen2011} including finite temperature properties \cite{Trotzky2010, McKay2015}. DBHM is an extension of this system, where disorder has been introduced to the underlying periodic potential resulting in an Hamiltonian with disorder \cite{Zhou2010, RayDissertation:2015}.

Effects of disorder on {\sl clean} systems have been discussed in general terms using scaling theory \cite{Harris1974,Chayes1986,Aharony1996,Wiseman1998}.  According to the Harris criterion \cite{Harris1974a}, if the condition $\nu d \ge 2$ is not satisfied, where $\nu$ is the critical exponent of the spatial correlation length ($\xi$) and $d$ is the spatial dimension, then the clean transition and the associated phases are sensitive to the presence of disorder, which can lead to a variety of system specific consequences \cite{Vojta2006}. For the Bose-Hubbard model in $d=3$ dimensions, the phase transition is in the (d+1)-XY model universality class that corresponds to mean-field like critical exponents $\nu = \frac{1}{2}$ \cite{Fisher1989}. This results in $\nu d = 1.5 < 2$ that does not satisfy the Harris criterion and leads to a fundamental change in the phase diagram of the DBHM. Consequently, there is a new disordered phase, the Bose-glass (BG) phase, that is an intermediary state between the disordered SF and the MI states (the MI may be absent contingent on the nature of the disorder) \cite{Fisher1989}. When the MI exists, the Theorem of Inclusions proves that a direct SF-MI phase transition is not possible and the BG must necessarily intervene \cite{Pollet2009}. The BG-MI is expected to be dominated by rare-regions effects similar to Griffiths type of transitions \cite{Kisker1997,Pollet2009,Krueger2011}. 

The BG is a unique disordered state that is composed of an insulating background embedded with puddles of SF \cite{Pollet2009,Meldgin2016}. As a result, it has the peculiar property of lacking long-range order while having infinite superfluid susceptibility and finite compressibility \cite{Fisher1989,Krauth1991,Scalettar1991,Pons2017}. The SF-MI transition of the BHM is replaced by the disordered SF-BG phase transition which, following very general arguments due to Chayes et al. \cite{Chayes1986}, has critical exponents that satisfy $\nu d \ge 2$. Recent work by Yao et al. has confirmed that $\nu = 0.88(5)$ \cite{Yao2014}. The SF-BG transition has been studied and confirmed via a synergistic study involving experiments with ultracold atomic gases and large scale QMC calculations \cite{Meldgin2016}. This transition is a peculiar gapless-SF to gapless-BG percolation-driven phase transition that nonetheless has all the qualities of a QPT \cite{Bissbort2010, Yao2014, Stauffer:1994}. The nature of the low-lying excitations of the two phases appears to be distinct: whereas for the SF they are non-localized sound modes \cite{LeggettBook2006}, the BG has localized excitations corresponding to the embedded SF puddles, though this requires further study. The properties of the disordered SF and the phase diagram of the DBHM are of key importance to our work. 

A central question with regards to the DBHM originates in the statistics of the disorder; it is typically of the static ``quenched'' kind; the disorder is site-dependent but fixed in time. In most studies, disorder is only considered on the local potential term of the Hamiltonian  \cite{Weinrib1983,Granato1986,Fisher1991,Fisher2004,Maestro2006, Pollet2009}, even though it may also be present in other terms of the Hamiltonian and may be pertinent in experiments \cite{Bissbort2010, Meldgin2016, Zhou2010, McKay2015, Pasienski2010}. In contrast to annealed disorder that can be handled by averaging the partition function, static disorder requires averaging the free energy over the different static realizations, making it much more technically challenging \cite{Vojta2013}. Furthermore, the perfect correlation in imaginary time means that the disorder cannot be integrated out in that direction so that it can have important consequences to the stability of the clean transition. Such instabilities can lead to new types of behavior such as the breakdown of self-averaging of observables \cite{Wiseman1995,Martino1998,Roy2006,Hegg2013,Dotsenko2017}. Additionally, it is unclear as to how the particular form of the disorder might affect the phase diagram. 

In this work we use Quantum Monte Carlo (QMC) to study the properties of the SF phase all the way up to the SF-BG interface of the DBHM at unit filling. We look at the qualitative and quantitative aspects of the SF in the presence of three different types of disorder: box, Gaussian and speckle. These are relevant to experiments with ultracold atomic gases \cite{White2009a, Pasienski2010, Meldgin2016} and may also apply to experiments with quasi-periodic lattices \cite{Fallani2007}. Following this, we study the statistics of the order parameters and consider finite size effects in the SF phase. We show that along the superfluid phase the order parameters are self-averaging, arguing that deviations from Gaussian behavior of probability distributions are mainly related to strong finite-size effects. 

In what follows, we discuss the details of the Disordered Bose-Hubbard model and our approach in studying it. Next, we discuss the effects of the disorder distribution on the phase diagram, following which we present disorder statistics for the superfluid fraction and the compressibility along the SF phase. The last set of results considers the probability distribution of the order parameters at specific points of the SF phase and details on finite size effects. We conclude with the relevance of our work as it applies to the DBHM, in general, and to experiments.


\section{Model and Method}
We simulate the model in a cubic lattice with $L$ sites in each direction using the Hamiltonian
\begin{equation}
\hat{H} = -t\sum_{<ij>} \hat{b}_i^\dagger \hat{b}_j + \frac{U}{2}\sum_i \hat{n}_i(\hat{n}_i - 1) + \sum_i (\epsilon_i - \mu) \hat{n}_i
\end{equation}
where $\hat{b}_i^\dagger$ ($\hat{b}_i$) is the bosonic creation (annihilation) operator on lattice site $i$, $t$ is the hopping amplitude to nearest-neighbors, $U$ is the interaction energy between pairs of Bosons on the same site, $\hat{n}_i$ is the number operator, $\mu$ is the chemical potential that controls the density and $\epsilon_i$ is the occupation energy for site $i$;  $\epsilon_i$ is sampled from a disorder distribution $P(\epsilon)$ with zero mean and standard deviation $\Delta$, which defines the strength of the disorder. For most of our work, we will consider a Gaussian disorder:
\begin{align}
P(\epsilon) = \frac{1}{\sqrt{2\pi}\Delta} \exp\left\{-\frac{\epsilon^2}{2\Delta^2}\right\}.
\label{eq:gaussian}
\end{align} 
In some cases, as we shall make explicit, we also consider a box distribution: 
\begin{align}
P(\epsilon) = \frac{1}{2\sqrt{3}\Delta}\Theta(\epsilon+\sqrt{3}\Delta)\Theta(-\epsilon+\sqrt{3}\Delta),
\label{eq:box}
\end{align}
where $\Theta(x)$ is the Heaviside step function. An exponential distribution corresponding to the speckle field used in ultracold atomic systems,
\begin{align}
P(\epsilon) = \frac{1}{\Delta} \exp\left\{-\frac{(\epsilon+\Delta)}{\Delta}\right\},
\label{eq:speckle}
\end{align}
is also considered in studying quantitative properties of the SF. For speckle systems, disorder is also present in the other terms of the Hamiltonian but has been found to have a small effect for the small to intermediate disorder strengths considered here \cite{RayDissertation:2015,Bissbort2010}.
Notice that in all cases the strength of the disorder is identical.

To perform the calculations we have used Stochastic Series Expansion (SSE), a powerful finite temperature exact method that samples the power series expansion of the partition function of lattice models using QMC techniques \cite{Sandvik1991,Sandvik1992}. The Monte Carlo sweeps involve a diagonal-update procedure that samples the order of the expansion and a directed-loop algorithm that samples the states over which the trace operation is performed \cite{Sandvik1999,Syljuaasen2002}. In the case of the DBHM, SSE works in the grand-canonical ensemble and employs the occupation-number basis set. The method has a world-line representation closely related to path-integral Monte Carlo (PIMC) methods \cite{Ceperley1995}. 

Although we use a finite temperature method, we consider low enough temperatures to obtain the ground state properties. The simplest estimate of the energy scale is set by the bandwidth for this 3D system -- equal to 12t. In ultra-cold atomic experiments -- systems of great interest since they can realize the DBHM -- the energy scale is set by the atomic recoil energy $E_R$: the kinetic energy imparted to an atom at rest by a photon from lasers used to setup the lattice \cite{White2009a, Meldgin2016}. In these terms, we define $\beta = E_R/k_B T$, where $k_B$ is the Boltzmann constant and $T$ is the temperature. Our simulations are done with $t / E_R = 1$ and $12 \beta t = 180$.

Of key interest to the DBHM are two observables: the compressibility,
\begin{equation}
\kappa \equiv \frac{\partial \langle N \rangle}{\partial \mu} = \beta \left[ \langle N^2 \rangle - \langle N \rangle^2 \right],
\end{equation}
where $N = \sum_i \hat{n}_i$ is the total number of particles, and the superfluid fraction,
\begin{equation}
\rho_s = \frac{m L^2}{\hbar^2 \beta N} \langle W^2 \rangle,
\end{equation}
where $m$ is the particle-mass and $W$ is the net winding-number \cite{Pollock1987}, which is easily computed since the method considers a decomposition of the lattice-Hamiltonian into bond-operators that link different lattice sites and we employ periodic boundary conditions. Thus, one just needs to check for operators that connect sites across the boundaries of the system. Fluctuations of $N$ that are necessary to calculate compressibilities are also readily achieved in the occupation-number basis. It is possible to identify each of the three phases of the model at zero temperature: the superfluid (SF), Bose-glass (BG) and Mott-insulator (MI), as described in Table \ref{tab:tab1}, using these two observables.
\begin{table}[t]
\caption{ \label{tab:tab1} Identification of the phases of the disordered Bose-Hubbard model in terms of order parameters.}
\begin{center}
\begin{tabular}{|c|c|c|}
\hline
Phase & Superfluid fraction & Compressibility \\
\hline
Superfluid (SF) & finite & finite  \\
Bose-glass (BG) & zero & finite  \\
Mott-insulator (MI) & zero & zero \\
\hline
\end{tabular}
\end{center}
\end{table}

Calculations were performed for unit filling since a commensurate filling is needed to see the emergence of the re-entrant superfluid (RSF) from the MI phase (see below). This implies that we need to find the appropriate $\mu$ for each disorder realization. The system is initialized in a random configuration then equilibrated before gathering properties of interest. It is then run long enough to reduce stochastic error. Specifically, we have taken care so that the order parameters have relative stochastic errors smaller than $2\%$. This enables us to study the effect of disorder averages independently of statistical errors due to sampling. 

Since we will be concerned with statistical properties of the SF in different parts of the phase diagram associated with the underlying properties of the disorder distribution, we have undertaken simulations for a large number of samples. The disorder average of an observable $X$ is denoted by $[X]$. Additionally, in order to get an insight into the scaling behavior, we have also undertaken studies over a range of system sizes. Specifically, we have considered linear lattice sizes $L = 6$, $8$, $10$, $12$, $16$ and $20$. 
To obtain the probability distribution of $X$ for a particular $L$, $\mathcal{P}_L (X)$, we consider a set of systems with different disorder realizations. Samples within a set are equivalent in the sense that they have exactly the same control parameters. Each sample in a set is independently simulated, giving us a single value of $X$. $\mathcal{P}_L (X)$ is then proportional to the histogram of these values. Also, since one of the specific properties of interest is the Gaussian-like nature of the different distributions, we employ a quantile-quantile measure that enables us to study the extent to which a distribution can be ascribed with Gaussian properties. The calculated quantities are sorted in ascending order according to their standard scores, $z_i^{data}$, while theoretical values are calculated from $z_i^{theo} = \Phi^{-1}\left[(i - 0.5) / n\right]$, where $n$ is the number of samples and $\Phi^{-1}$ is the standard normal quantile function. These values are then plotted against each other; the points will lie on a straight line of unit slope if the statistics are Gaussian.

\section{Results and Discussion}

\subsection{Phase diagram}

\begin{figure}[t]
\begin{center}
\includegraphics{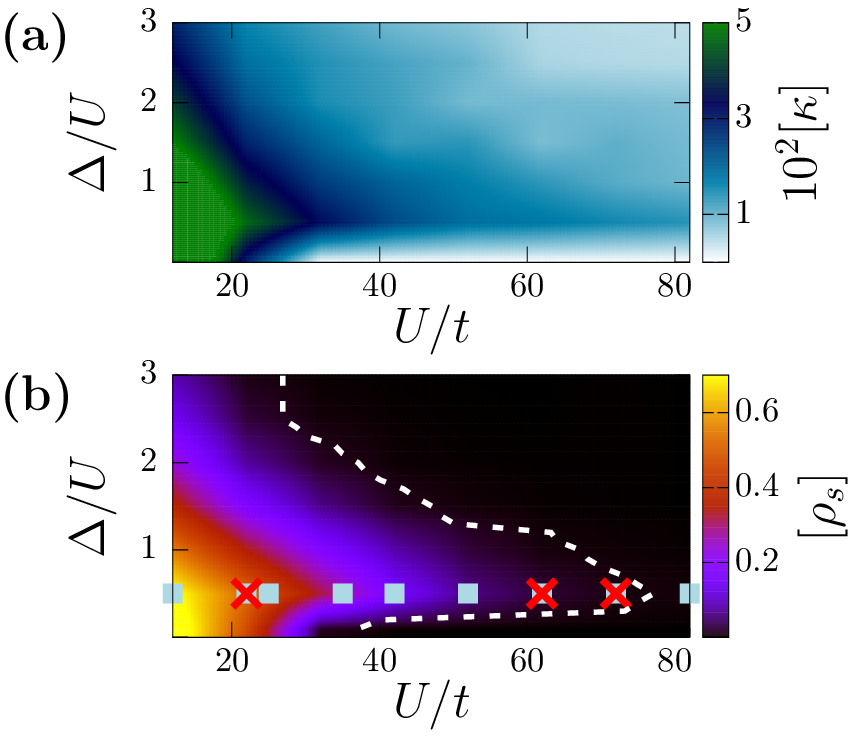}
\end{center}
\caption{ \label{fig:fig1} Order parameters (color scale) of the diagonal-disordered Bose-Hubbard model at unit filling for a $L = 6$ cubic lattice as a function of  $\Delta/U$  (vertical axis) and $U/t$ (horizontal axis). {\bf (a)} Compressibility per particle $[\kappa]$. {\bf (b)} Superfluid fraction $[\rho_s]$. The light-blue squares indicate points with $\Delta / U = 0.5$ where we have performed analysis of the disorder statistics (see Fig. \ref{fig:fig3}). The red crosses refer to the systems in Figs. (\ref{fig:fig4}-\ref{fig:fig7}). The white-dashed line indicates the approximate location of the SF-BG phase boundary that is obtained by delimitating $\rho_s < 1/6^3$.}
\end{figure}

Fig. \ref{fig:fig1} shows the order parameters obtained after averaging over 40 disorder realizations for a $L = 6$ lattice. We note that this phase diagram, constructed for Gaussian disorder, bears a striking resemblance to those obtained for disorder generated from speckle-systems used in optical lattices where $P(\epsilon)$ is given by the exponential distribution of Eq. \ref{eq:speckle} \cite{RayDissertation:2015}. It also bears the same features seen for the box disorder of Eq. \ref{eq:box} \cite{Gurarie2009}. Here, we employ the normalization $\Delta / U$ because the physics of the system is influenced by the local energy scales that arises due the interplay of the occupation energies $\epsilon_i$ and the interaction energy $U$. Specifically, local shifts in $\epsilon_i$ due to disorder can decrease (increase) the gap associated with multi-particle occupation thereby facilitating (hindering) delocalization (localization). This, in-turn, affects the formation of a global SF. These effects are illustrated in the obtained phase diagram.

The most remarkable feature arising from disorder
is the resurgence of superfluidity when the clean system would be insulating -- the so called re-entrant superfluid phase (RSF) \cite{Pollet2009,RayDissertation:2015}. It can be noticed in Fig. \ref{fig:fig1} by the extension of finite superfluid fractions to regions where it is zero in the clean system that has a critical point at $U/t = 29.34(2)$ \cite{Capogrosso-Sansone2007,Lacki2016}. 
The RSF typically arises as a finger and appears to be controlled entirely by the disorder strength $\Delta$: this shape and qualitative aspects arise in all three types of disorder considered. The superfluidity arises from a common percolation mechanism \cite{Yao2014}. 

Typically, the destruction of the MI requires the gap to close locally. This mechanism is responsible for the creation of local SF puddles that are ubiquitous in the BG phase \cite{Meldgin2016}. For the additional requirement of globally coherent superflow, the puddles must be connected over the disorder terrain and delocalization must not be too energetically prohibitive. For a given interaction strength and weak disorder, though the creation of SF puddles is possible (for unbounded disordered system this will be the case no matter how small the disorder), 
they are too rare for achieving global superfluidity. Additionally, with increasing $U/t$, delocalization is penalized due to energetics and, consequently, the creation of SF puddles is suppressed. These effects explain the behavior of the superfluid fraction for $\Delta/U<0.5$ and $U/t>29.34(2)$. For intermediate disorder strengths, the puddles proliferate and the particles are able to tunnel through the disorder terrain across different puddles, thereby leading to a globally coherent superflow. Thus, the RSF extends to large values of $U/t$, until the energetic cost of delocalization is too large to support a superflow. For larger disorder strengths, there are patches of space where, relative to the chemical potential of the system, the disorder is so large that it creates barriers in the form of hills or valleys that the particles cannot traverse, resulting in the loss of global coherence. For increasing disorder, these patches proliferate. The net effect of these tendencies is the resulting RSF finger. To first order then, the disorder strength plays the dominant role in describing this aspect. 

\begin{figure}[t]
\begin{center}
\includegraphics{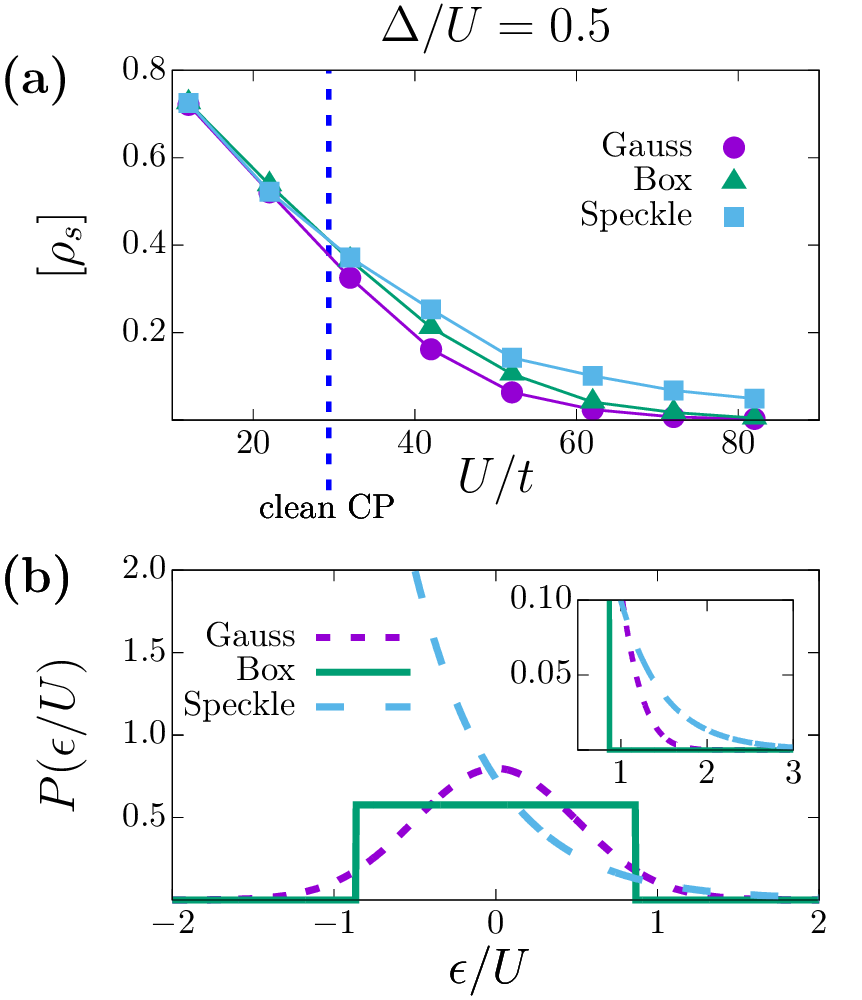}
\end{center}
\caption{ \label{fig:fig2} {\bf(a)} Superfluid fraction as a function of $U/t$ for different types of disorder distribution at $\Delta/U = 0.5$. Error bars are too small to be seen. {\bf(b)} Disorder distributions considered in calculating the superfluid fraction (Eqs. \ref{eq:gaussian} to \ref{eq:speckle}). Inset: Zoom on the tails of the distributions. See text for discussion.}
\end{figure}

The precise disorder distribution is less important, at least for the disorder types and strengths considered here. Notice that negative shifts relative to the chemical potential result in deep wells that can have two different effects. Intermediate values of $\epsilon$ (compared to the gap $\sim U/2$) can help in lowering the energy cost associated with the multi-particle occupation leading to delocalization of particles. However, for too large negative shifts, the wells may be so deep as to localize particles. Positive shifts, on the other hand, increase the cost associated with site occupation and serve to prevent hopping and delocalization. From Fig. \ref{fig:fig2} (a), it is evident that for large superfluid fraction ($\rho_s$) there is very little difference as the SF is able to screen the disordered potential. For smaller values, corresponding to an increase in $U/t$, the  effects of disorder become pronounced and the differences in the distributions lead to quantitative differences in the superfluid order parameter. According to the distributions shown in Fig. \ref{fig:fig2} (b), $\sim 21.13 \%$ of the sites of the box disorder have $\epsilon > \Delta$ compared to $\sim 15.87\%$ for Gaussian and $\sim 13.53\%$ for speckle distributions. Conversely, for $\epsilon < -\Delta$, box and Gaussian disorders have the same numbers as before but the exponential distribution has no such sites. This means that, in the exponential case, only the positive tail of the distribution contributes to the localization effects, whereas the negative side facilitates in delocalization. It is then unsurprising that SF is enhanced for the speckle relative to the other types of disorder. The differences between box and Gaussian disorders is due to the interplay of 
{\bf(i)} the number and distribution of sites that have sufficiently large (negative) disorder to create SF puddles, and 
{\bf(ii)} the number and distribution of sites that create patches of impassible terrain either due to large positive or negative shifts relative to the chemical potential. It appears to be the case that the SF in this regime of parameters is enhanced more by delocalization effects due to (i) than it is suppressed by localization effects due to (ii). 

In this paper we are concerned with the disorder related statistical properties of the SF all the way to the SF-BG transition. In the past, statistical properties of the SF-BG transition and the SF phase in the presence of disorder have been speculated on but, to the best of our knowledge, not explicitly studied in 3D systems. Experiments with ultracold atomic gases in disordered lattices typically find very small changes in condensate fraction values using time-averaged measurements. The time averaging is a proxy for disorder averaging, since the focus of the laser used to produce speckle patterns typically changes in time leading to different disorder distributions \cite{Meldgin2016, RayDissertation:2015}. 

\subsection{Disorder statistics along the superfluid phase}
A central quantity in the investigation of disorder statistics is the {\sl relative variance} of an observable $X$ that quantifies some physical property, defined as
\begin{align}
\mathcal{D}_X (L) = (\Delta X)^2 / [X]^2,
\label{eq:relvar}
\end{align}
where $(\Delta X)^2$ is the disorder-variance of $X$. The scaling of $\mathcal{D}$ to the thermodynamic limit is of paramount importance. If $\mathcal{D}_X (L) \to 0$ when $L \to \infty$, $X$ is said to be a {\sl self-averaging} property. More specifically, for $\mathcal{D}_X (L) \sim 1 / L^r$, if $r = d$, the spatial dimension of the system, $X$ exhibits {\sl strong} self-averaging, whereas {\sl weak} self-averaging is the case when $r < d$ \cite{Roy2006}. Either way, a single disorder realization in the thermodynamic limit is sufficient to capture all the physics of the disordered system related to $X$. Conversely, when $\mathcal{D}_X (L)$ does not vanish with increasing system sizes, $X$ is said to be a {\sl non-self-averaging} property and it is unclear whether it is possible to assign universal behavior in such disordered systems \cite{Aharony1996}.\\ 

\begin{figure}[t]
\begin{center}
\includegraphics{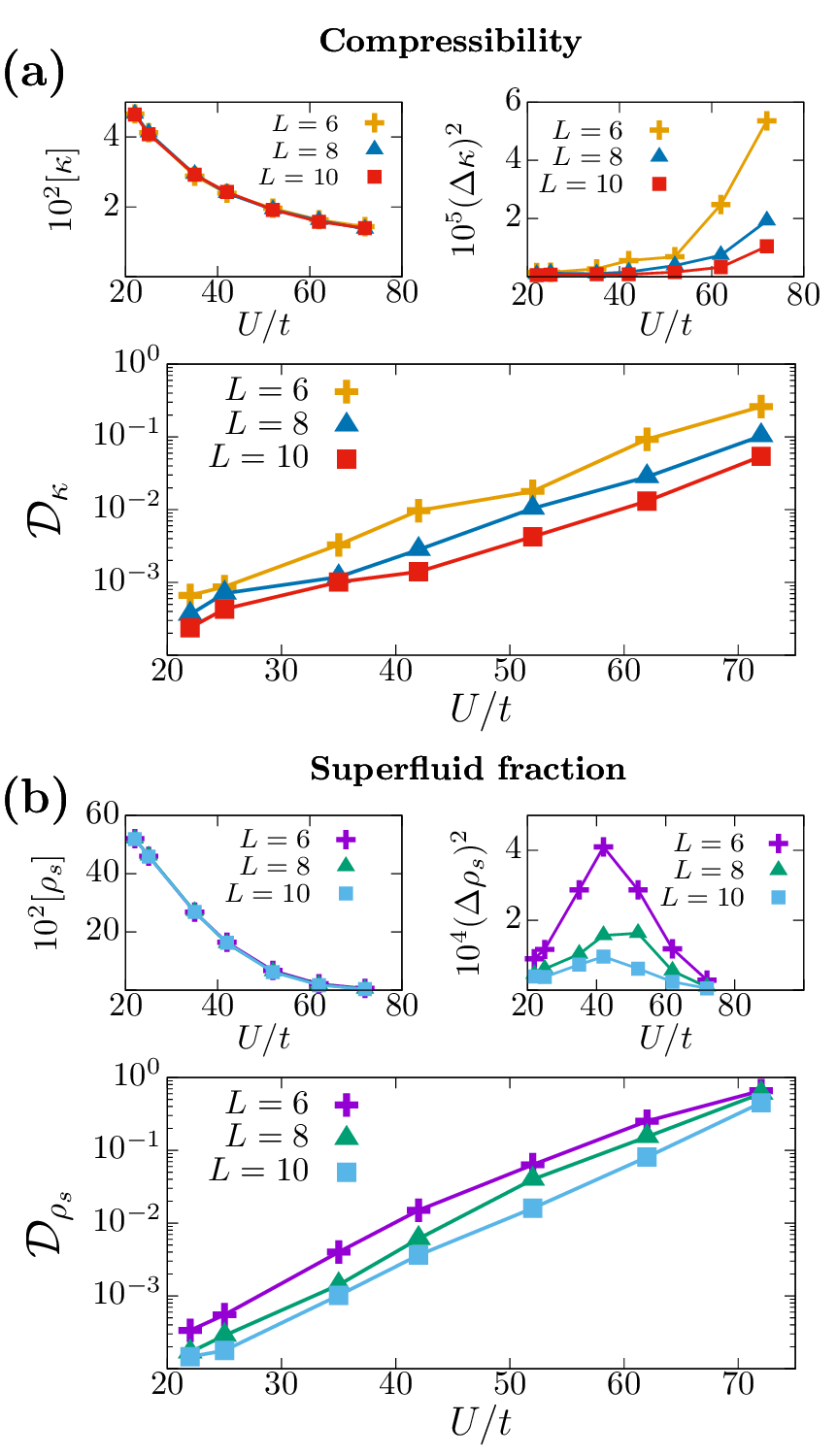}
\end{center}
\caption{ \label{fig:fig3} Disorder-statistical quantities of the order parameters along the line $\Delta / U = 0.5$ as a function of $U/t$ for lattice sizes $L = 6$, $8$ and $10$. {\bf (a)} Compressibility per particle average (top-left), variance (top-right) and relative variance (logarithmic scale, bottom). Lines are guides to the eye. {\bf (b)} Same quantities for the superfluid fraction. Error bars are too small to be seen. See text for discussion.}
\end{figure}

Fig. \ref{fig:fig3} shows the relative variances of the superfluid fraction $\rho_s$ and compressibility per particle $\kappa$ for $\Delta/U = 0.5$ at the points marked by white squares in Fig. \ref{fig:fig1}. These points traverse the interior of the SF phase, starting from the regular SF, through the SF-MI transition point of the clean system, and terminating in the RSF part of the phase diagram. We note that throughout this range, $\mathcal{D}$ decreases for increasing lattice sizes as is expected from general renormalization group arguments. This points to the existence of an attractive zero-disorder fixed point dictating the behavior of the system at large length scales \cite{Vojta2014,Vojta2016}. The superfluid phase is, therefore, expected to have self-averaging properties. However, it may still have strong finite-size effects for small to intermediate systems sizes worth exploring since such systems are frequently studied in experiments \cite{White2009a, Pasienski2010, Meldgin2016, Fallani2007}. Towards this end, we also notice that $\mathcal{D}_{\rho_s,\kappa}$ is monotonically increasing with $U/t$,  indicating that the fluctuations from sample to sample of both $\rho_s$ and $\kappa$ become larger compared to their averages -- they grow by four orders of magnitude in this range! The standard-deviation associated with disorder averaging thus grows from a few percent, which is within the size of the statistical error in every sample, to about a hundred percent. The basic mechanism follows from our discussion earlier that an increased $U/t$ leads to a reduction in the number and uniformity in the distribution of the SF puddles. Consequently, only a few puddles are able to participate and maintain the global superflow. A reduction in the number of superfluid channels makes the total flow more susceptible to the specificities of the distribution of the disorder potential. Basically, larger quantities of superfluid are able to screen the effects of the disorder distribution more effectively leading to a reduction in $\mathcal{D}_{\rho_s,\kappa}$. 

\begin{figure}[t]
\begin{center}
\includegraphics{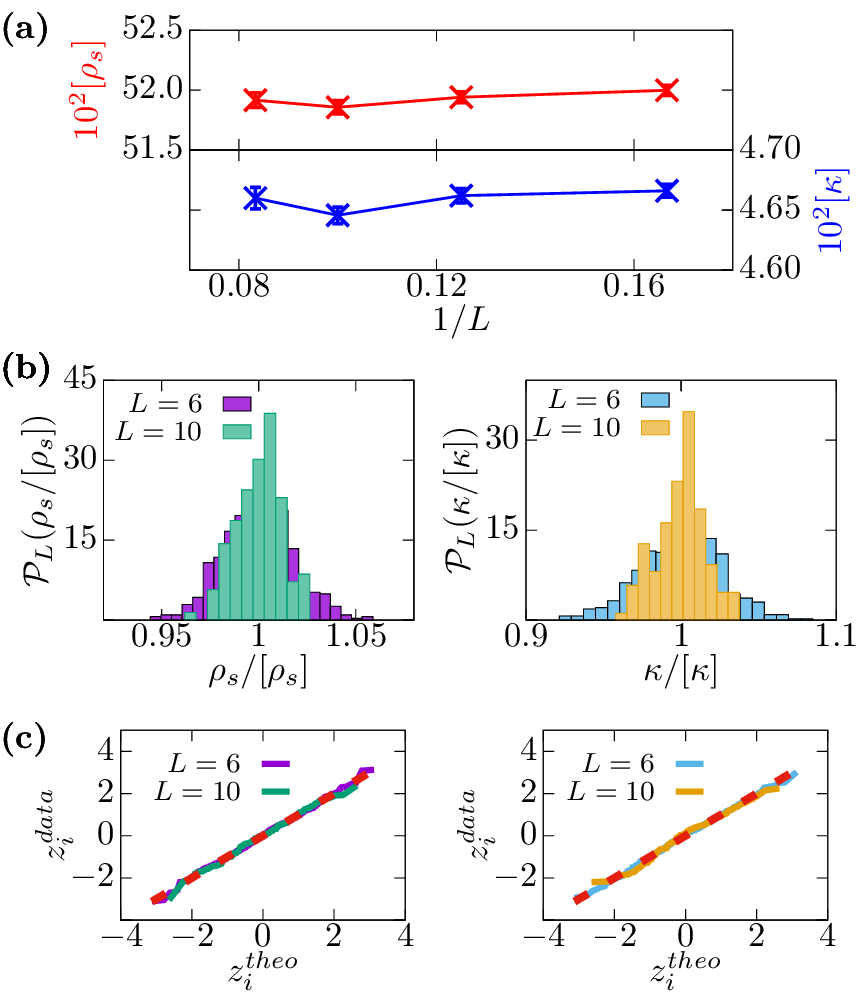}
\end{center}
\caption{ \label{fig:fig4} Disorder statistics of the order parameters at ($U/t = 22.0$, $\Delta / U = 0.5$). {\bf (a)} Disorder averages of the superfluid fraction $\rho_s$ (red) and compressibility per particle $\kappa$ (blue) as a function of inverse linear lattice size. Lines are guides to the eye. Lattice sizes used were $L = 6 [556]$, $8 [235]$, $10 [120]$, and $12 [70]$, where $[...]$ indicates the number of samples. {\bf (b)} Histograms of the values for $\rho_s$ (left) and $\kappa$ (right) relative to their respective averages for lattice sizes $L = 6$ and $10$. {\bf (c)} Normal quantile-quantile plots of the associated distributions, the red-dashed line indicates perfect normal behavior.}
\end{figure}

Another interesting feature that is shown in Fig. \ref{fig:fig3} is the peak in the variance of the superfluid order parameter, $(\Delta \rho_s)^2$. This is a finite size effect arising from the way SF domains transform to MI (and vice-versa) across the SF-MI transition point in the clean system. For small systems the formation of a globally connected SF puddle is highly susceptible to the underlying disorder distribution. As $U/t \rightarrow (U/t)_c$ of the clean transition, the SF supporting domains, which are few in small systems, transform to MI, undergoing the usual route of critical fluctuations. This crossover behavior is eliminated as the system size is increased. Its effect on the compressibility is weaker because even when global SF is absent there are still disconnected SF puddles. (Note that at $\Delta/U = 0.5$, for both bounded and unbounded disorder, there is always a finite probability for a collection of sites to locally close the MI gap leading to formation of puddles of SF.) Such puddles contribute locally to a finite value of compressibility and together with zero compressibility domains due the MI lead to a wider range of behavior to average over for $\kappa$ compared to $\rho_s$. As a result, the variance for the compressibility, $(\Delta \kappa)^2$, is simply a monotonically increasing function of $U/t$.

\subsection{Disorder distribution of order parameters}
Next we consider the finite size scaling and the histograms associated with the order parameters of interest. We have undertaken a range of simulations to prepare ensembles of distributions for different system sizes. This is done in a way to ensure that the disorder distribution $P(\epsilon)$ is sampled evenly for all ensembles. To start with, consider the distributions $\mathcal{P}_L(\rho_s)$ and $\mathcal{P}_L(\kappa)$ for ($U/t = 22.0$, $\Delta / U = 0.5$) shown in Fig. \ref{fig:fig4}. Deep within the SF phase, the global superflow is so large that it basically screens the disorder potential very effectively. This leads to very small finite-size errors. The SF channels are not susceptible to the variations of the disorder potential between samples. As expected, the finite-size scaling goes from a broader Gaussian distribution at $L = 6$ to a narrow distribution for $L = 10$. This behavior continues for interior points in the SF and RSF where $\rho_s$ is large enough to screen the disorder potential.

\begin{figure}[t]
\begin{center}
\includegraphics{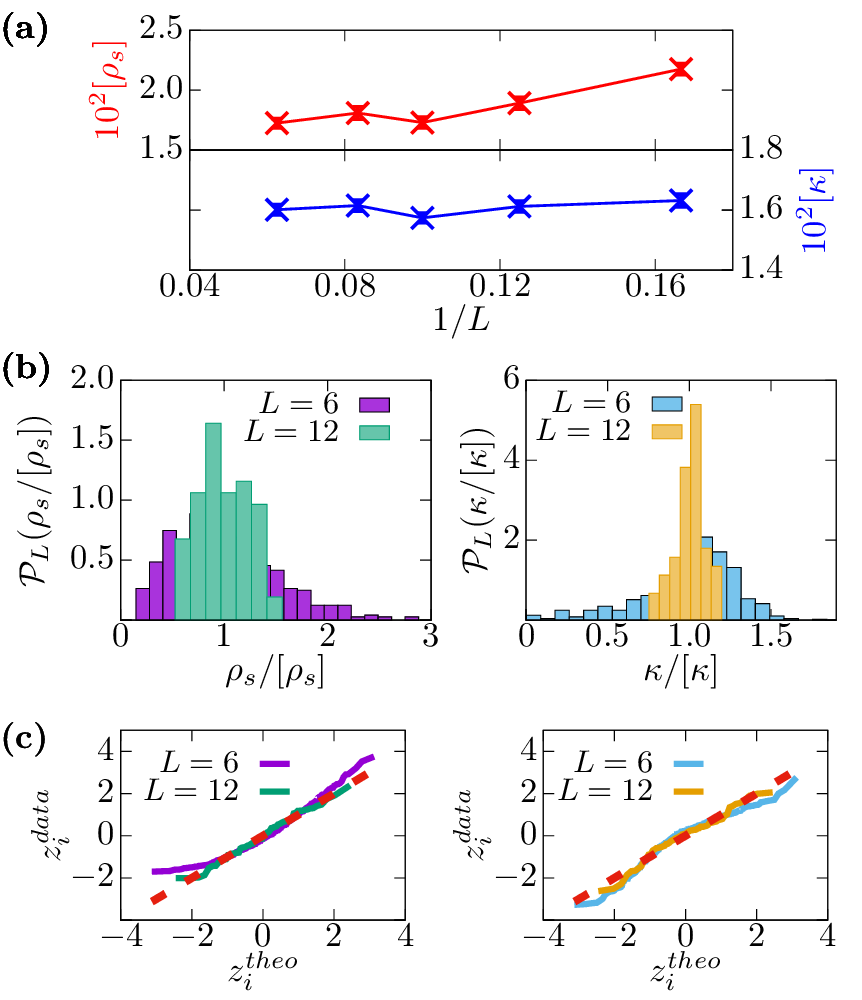}
\end{center}
\caption{ \label{fig:fig5} Disorder statistics of the order parameters at ($U/t = 62.0$, $\Delta / U = 0.5$). {\bf (a)} Same quantities as in Fig. \ref{fig:fig4}. Lattice sizes used were $L = 6[556]$, $8[235]$, $10[120]$, $12[70]$ and $16[29]$. {\bf (b)} Histograms for lattice sizes $L = 6$ and $12$. {\bf (c)} Associated normal quantile-quantile plots.}
\end{figure}

More interesting features start arising near the SF-BG boundary, where $\rho_s$ is small. Consider first the point at ($U/t = 62.0$, $\Delta / U = 0.5$), shown in Fig. \ref{fig:fig5}. For the smallest lattice size $L = 6$ we notice a significant skewness in both $\mathcal{P}_L(\rho_s)$ and $\mathcal{P}_L(\kappa)$. There are a significant fraction of samples with low $\rho_s$ and $\kappa$ values, pointing to both the low number of SF puddles and to the lack of enough hopping events that lead to global coherence. The larger than average valued tail associated to large $\rho_s$ corresponds to aberrantly larger number of SF puddles or an increase of connectivity. In principle, it is possible to distinguish between the two by considering the condensate fraction associated with the different puddles, but it is numerically challenging owing to statistical noise \cite{RayDissertation:2015, Meldgin2016}. In any case, once the system size is increased, the distribution of puddles reaches typical trends and the histograms become Gaussian as expected. 

\begin{figure}[t]
\begin{center}
\includegraphics{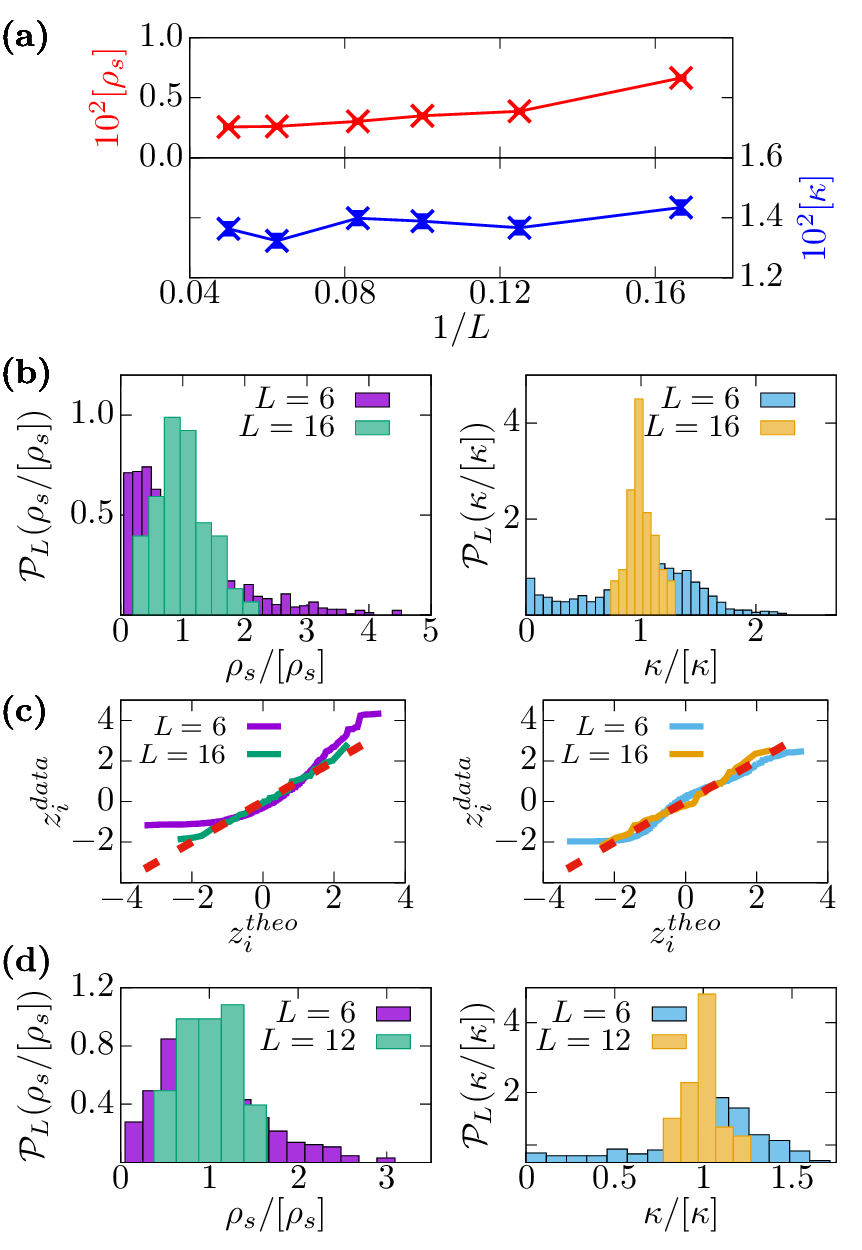}
\end{center}
\caption{ \label{fig:fig6} Disorder statistics of the order parameters at ($U/t = 72.0$, $\Delta / U = 0.5$). {\bf (a)} Same quantities as in Figs. \ref{fig:fig4}, \ref{fig:fig5}. Lattice sizes used were $L = 6[1250]$, $8[528]$, $10[270]$, $12[157]$, $16[66]$ and $20[34]$. {\bf (b)} Histograms  for lattice sizes $L = 6$ and $16$. {\bf (c)} Associated normal quantile-quantile plots. {\bf (d)} Corresponding histograms for simulations done with box-disorder.}
\end{figure}

The final point we consider is located at ($U/t = 72.0$, $\Delta / U = 0.5$), close to the SF-BG boundary, where the relative variance of the order parameters is the largest. This is a rather tenuous point in the RSF part of the phase diagram. It is clear that increasing or decreasing the disorder strength would lead to a loss of global coherence. The percolating SF clusters and insulating disorder clusters are so arranged that they barely support global coherence -- $\rho_s$ is only about $0.2\%$. Results have been presented in Fig. \ref{fig:fig6}. The disorder averages clearly exhibit very strong finite-size effects. For $\rho_s$, these effects persist up to lattices of size $L \sim 16$, beyond which the values agree within the error bars. The same effect is visible in $\kappa$ but it is not as strong, persisting only up to $L \sim 8$. This behavior is evident in the shape of the probability distributions that are highly skewed and broad for small lattice sizes. In the case of $\rho_s$, there is a range of behavior for small lattices with samples at one end of the distribution having as much as four times larger values than the average, whereas for the other end, samples have almost no superfluid at all. 

One of the remarkable features of the histograms associated with the small system sizes $L = 6$ is that a considerable number of samples appear to be entirely insulating with no SF present in them. Additionally, a large fraction of the samples appear to also be significantly skewed towards having lower than average $\rho_s$ values. Contrasting this behavior against histograms generated from box-disorder where this additional skewness is less pronounced illustrates the subtle way in which the shape of the disorder distribution affects the statistics of the SF puddles. As we have mentioned earlier, it appears that the SF is favored more by the delocalization effects due to the number of sites that can close the gap ($21.13\%$ for box vs. $15.87\%$ for Gaussian). 

However, as we consider larger systems, the distributions consistently start narrowing and non-Gaussian features are mitigated as both the shape of the histograms and the normal quantile-quantile plots indicate. This reduction of the relative variance $\mathcal{D}(L)$ therefore shows that both order parameters are self-averaging properties of the model even when we approach the phase transition.

\subsection{Scaling of relative variances}
\begin{figure}[t]
\begin{center}
\includegraphics{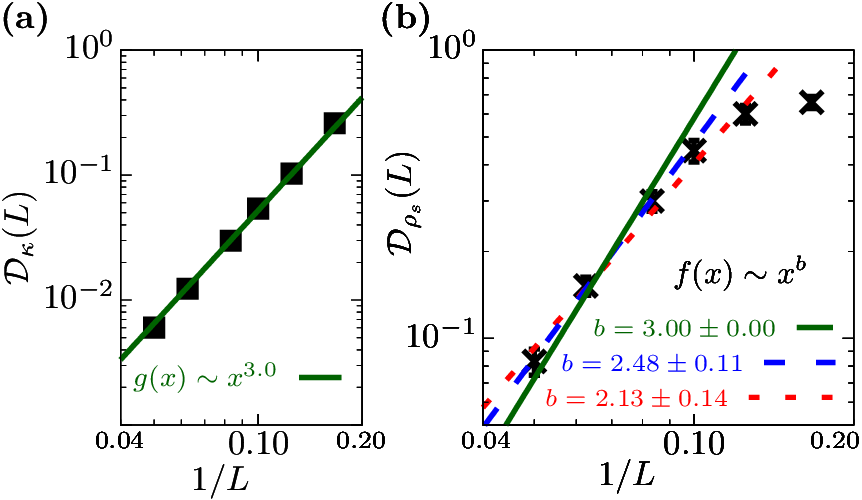}
\end{center}
\caption{ \label{fig:fig7} Relative variance of the order parameters (log scale) as function of inverse linear lattice size (log scale) for ($U/t = 72.0$, $\Delta / U = 0.5$). {\bf (a)} Compressibility per particle. The dark-green line is a fit to a $3.0$ power law. {\bf (b)} For the superfluid fraction, a power law was fitted excluding the smallest lattice size $L = 6$ (dashed-red), and also the three smallest $L = 6$, $8$ and $10$ (dashed-blue) in order to capture the behavior in the thermodynamic limit. The solid dark-green line is a fit to a $3.0$ power law. See text for discussion.}
\end{figure}

To address the question of whether there is {\sl strong} or {\sl weak} self-averaging, we made more precise estimates of the scaling of the relative variance that are shown in Fig. \ref{fig:fig7}. We fit the relative variance to $f(x) = x^b$ to estimate the scaling coefficient $b$. Recall from our earlier discussion that a strong self-averaging implies the relative variance must scale according to the dimension $d$ of the system.

It is evident that the compressibility exhibits strong self-averaging behavior, since its relative variance is undoubtedly scaling with a $d = 3$ power law, as indicated by the dark-green line that falls right on the calculated points. 
The situation for the superfluid order parameter is more complicated. Although the general trend  shows a decrease in the relative variance with increasing lattice size, we are unable to scale to very large values of L and perform sufficient disorder averaging in order to confidently estimate the scaling exponent. To get some idea of what the scaling might be, we fit the data to two different data sets that disregard small lattice sizes in order to reduce the bias associated with large non-Gaussian behavior.
The red dashed line in Fig. \ref{fig:fig7} is obtained by excluding $L = 6$ points and the corresponding fit is $b = 2.13 \pm 0.14$. We also fit it to the blue dashed line by additionally excluding $L = 8$, $10$ lattice sizes to obtain $b = 2.48 \pm 0.11$. For larger values of $L$ it might very well be the case that $b \rightarrow 3$, thereby suggesting there is strong self-averaging even this close to the SF-BG boundary. On the other hand, for practical purposes, even a weak self-averaging is remarkable, given that the critical point cannot be self-averaging in any finite system.

\section{Conclusions}
Our study of the disorder statistics of the SF phase points, beyond any doubt, to the existence of self-averaging of $\kappa$ and $\rho_s$. This property extends throughout the SF phase all the way up to the SF-BG boundary. The self-averaging is of the strong type for $\kappa$. Although we have not been able to conclude the exact value of the scaling for $\rho_s$, it appears to be at least weakly self-averaging.  As a consequence we expect that most experiments with ultra-cold atomic gases can safely report observable values without being concerned about disorder averaging. Even near the SF-BG interface we suspect that the statistical and systematic errors associated with imaging and time-of-flight based measurements will be much larger than disorder averaging related errors. However, finite-size errors might still be significant especially when system sizes are small. Our results are directly applicable to different types of disorder experiments, as we have studied effects of unbounded Gaussian type of disorder that is related to the exponential type of disorder of speckle-fields \cite{White2009a}, as well as the more idealistic box type of disorder that can be realized in experiments with homogeneous traps  \cite{Lopes2017,Mukherjee2017}. 

Although we have only reported data for $\rho_s$ and $\kappa$, we have also studied other physical quantities of the system. In particular, the energy exhibits strong self-averaging. The condensate fraction ($n_0$) exhibits features similar to those of the superfluid fraction, which indicates that it is at least weakly self-averaging, but once again we were unable to conclude the exact value of the scaling of relative variances for this case, which is even more subtle because it is computationally challenging to reduce the stastistical errors that come from the sampling procedure, since $n_0$ results from the diagonalization of the single-particle density matrix of the system -- a non-linear operation. We have also considered larger values of the disorder strength $\Delta / U$ that are in complete agreement with the conclusions presented here. 

The results we have presented corroborates our understanding of the percolation mechanism describing different aspects of the phase diagram. We were able to show that the strength of the disorder distribution plays the dominant role in describing the qualitative aspects of the phase diagram. The shapes of the distribution come into effect when quantitative comparisons are concerned and also with regards to finite-size effects. In the future, it would be interesting to see if our analysis can be extended to the Bose-glass phase in order to characterize when non-self-averaging behavior sets in. There are also interesting prospects with regards to studying the percolation problem in these quantum systems as well. Particularly, owing to the tunneling type of phenomena that governs the connection of percolating clusters, there might be significant differences in the fractal properties of the transition when compared against the standard classical picture \cite{Stauffer:1994}.

\begin{acknowledgements}
This work was supported by the Conselho Nacional de Desenvolvimento Científico e Tecnológico (CNPq) under grants nos. 141242/2014-0 and 232682/2014-3 (Science Without Borders program) and by the Fundação de Amparo à Pesquisa do Estado de São Paulo (FAPESP). Part of the computations were performed at the Centro Nacional de Processamento de Alto Desempenho em São Paulo (CENAPAD-SP), a high-performance computing facility at Unicamp. All simulations were done using CSSER, a parallel SSE based library available at \url{https://github.com/ushnishray/CSSER}.
\end{acknowledgements}



\begin{thebibliography}{66}%
\makeatletter
\providecommand \@ifxundefined [1]{%
 \@ifx{#1\undefined}
}%
\providecommand \@ifnum [1]{%
 \ifnum #1\expandafter \@firstoftwo
 \else \expandafter \@secondoftwo
 \fi
}%
\providecommand \@ifx [1]{%
 \ifx #1\expandafter \@firstoftwo
 \else \expandafter \@secondoftwo
 \fi
}%
\providecommand \natexlab [1]{#1}%
\providecommand \enquote  [1]{``#1''}%
\providecommand \bibnamefont  [1]{#1}%
\providecommand \bibfnamefont [1]{#1}%
\providecommand \citenamefont [1]{#1}%
\providecommand \href@noop [0]{\@secondoftwo}%
\providecommand \href [0]{\begingroup \@sanitize@url \@href}%
\providecommand \@href[1]{\@@startlink{#1}\@@href}%
\providecommand \@@href[1]{\endgroup#1\@@endlink}%
\providecommand \@sanitize@url [0]{\catcode `\\12\catcode `\$12\catcode
  `\&12\catcode `\#12\catcode `\^12\catcode `\_12\catcode `\%12\relax}%
\providecommand \@@startlink[1]{}%
\providecommand \@@endlink[0]{}%
\providecommand \url  [0]{\begingroup\@sanitize@url \@url }%
\providecommand \@url [1]{\endgroup\@href {#1}{\urlprefix }}%
\providecommand \urlprefix  [0]{URL }%
\providecommand \Eprint [0]{\href }%
\providecommand \doibase [0]{http://dx.doi.org/}%
\providecommand \selectlanguage [0]{\@gobble}%
\providecommand \bibinfo  [0]{\@secondoftwo}%
\providecommand \bibfield  [0]{\@secondoftwo}%
\providecommand \translation [1]{[#1]}%
\providecommand \BibitemOpen [0]{}%
\providecommand \bibitemStop [0]{}%
\providecommand \bibitemNoStop [0]{.\EOS\space}%
\providecommand \EOS [0]{\spacefactor3000\relax}%
\providecommand \BibitemShut  [1]{\csname bibitem#1\endcsname}%
\let\auto@bib@innerbib\@empty
\bibitem [{\citenamefont {Fisher}\ \emph {et~al.}(1989)\citenamefont {Fisher},
  \citenamefont {Weichman}, \citenamefont {Grinstein},\ and\ \citenamefont
  {Fisher}}]{Fisher1989}%
  \BibitemOpen
  \bibfield  {author} {\bibinfo {author} {\bibfnamefont {M.~P.~A.}\
  \bibnamefont {Fisher}}, \bibinfo {author} {\bibfnamefont {P.~B.}\
  \bibnamefont {Weichman}}, \bibinfo {author} {\bibfnamefont {G.}~\bibnamefont
  {Grinstein}}, \ and\ \bibinfo {author} {\bibfnamefont {D.~S.}\ \bibnamefont
  {Fisher}},\ }\href {\doibase 10.1103/physrevb.40.546} {\bibfield  {journal}
  {\bibinfo  {journal} {Physical Review B}\ }\textbf {\bibinfo {volume} {40}},\
  \bibinfo {pages} {546} (\bibinfo {year} {1989})}\BibitemShut {NoStop}%
\bibitem [{\citenamefont {Ma}\ \emph {et~al.}(1986)\citenamefont {Ma},
  \citenamefont {Halperin},\ and\ \citenamefont {Lee}}]{Ma1986}%
  \BibitemOpen
  \bibfield  {author} {\bibinfo {author} {\bibfnamefont {M.}~\bibnamefont
  {Ma}}, \bibinfo {author} {\bibfnamefont {B.~I.}\ \bibnamefont {Halperin}}, \
  and\ \bibinfo {author} {\bibfnamefont {P.~A.}\ \bibnamefont {Lee}},\ }\href
  {\doibase 10.1103/physrevb.34.3136} {\bibfield  {journal} {\bibinfo
  {journal} {Physical Review B}\ }\textbf {\bibinfo {volume} {34}},\ \bibinfo
  {pages} {3136} (\bibinfo {year} {1986})}\BibitemShut {NoStop}%
\bibitem [{\citenamefont {Krauth}\ \emph {et~al.}(1991)\citenamefont {Krauth},
  \citenamefont {Trivedi},\ and\ \citenamefont {Ceperley}}]{Krauth1991}%
  \BibitemOpen
  \bibfield  {author} {\bibinfo {author} {\bibfnamefont {W.}~\bibnamefont
  {Krauth}}, \bibinfo {author} {\bibfnamefont {N.}~\bibnamefont {Trivedi}}, \
  and\ \bibinfo {author} {\bibfnamefont {D.}~\bibnamefont {Ceperley}},\ }\href
  {\doibase 10.1103/physrevlett.67.2307} {\bibfield  {journal} {\bibinfo
  {journal} {Physical Review Letters}\ }\textbf {\bibinfo {volume} {67}},\
  \bibinfo {pages} {2307} (\bibinfo {year} {1991})}\BibitemShut {NoStop}%
\bibitem [{\citenamefont {Scalettar}\ \emph {et~al.}(1991)\citenamefont
  {Scalettar}, \citenamefont {Batrouni},\ and\ \citenamefont
  {Zimanyi}}]{Scalettar1991}%
  \BibitemOpen
  \bibfield  {author} {\bibinfo {author} {\bibfnamefont {R.~T.}\ \bibnamefont
  {Scalettar}}, \bibinfo {author} {\bibfnamefont {G.~G.}\ \bibnamefont
  {Batrouni}}, \ and\ \bibinfo {author} {\bibfnamefont {G.~T.}\ \bibnamefont
  {Zimanyi}},\ }\href {\doibase 10.1103/physrevlett.66.3144} {\bibfield
  {journal} {\bibinfo  {journal} {Physical Review Letters}\ }\textbf {\bibinfo
  {volume} {66}},\ \bibinfo {pages} {3144} (\bibinfo {year}
  {1991})}\BibitemShut {NoStop}%
\bibitem [{\citenamefont {Pollet}\ \emph {et~al.}(2009)\citenamefont {Pollet},
  \citenamefont {Prokof'ev}, \citenamefont {Svistunov},\ and\ \citenamefont
  {Troyer}}]{Pollet2009}%
  \BibitemOpen
  \bibfield  {author} {\bibinfo {author} {\bibfnamefont {L.}~\bibnamefont
  {Pollet}}, \bibinfo {author} {\bibfnamefont {N.~V.}\ \bibnamefont
  {Prokof'ev}}, \bibinfo {author} {\bibfnamefont {B.~V.}\ \bibnamefont
  {Svistunov}}, \ and\ \bibinfo {author} {\bibfnamefont {M.}~\bibnamefont
  {Troyer}},\ }\href {\doibase 10.1103/physrevlett.103.140402} {\bibfield
  {journal} {\bibinfo  {journal} {Physical Review Letters}\ }\textbf {\bibinfo
  {volume} {103}} (\bibinfo {year} {2009}),\
  10.1103/physrevlett.103.140402}\BibitemShut {NoStop}%
\bibitem [{\citenamefont {White}\ \emph {et~al.}(2009)\citenamefont {White},
  \citenamefont {Pasienski}, \citenamefont {McKay}, \citenamefont {Zhou},
  \citenamefont {Ceperley},\ and\ \citenamefont {DeMarco}}]{White2009a}%
  \BibitemOpen
  \bibfield  {author} {\bibinfo {author} {\bibfnamefont {M.}~\bibnamefont
  {White}}, \bibinfo {author} {\bibfnamefont {M.}~\bibnamefont {Pasienski}},
  \bibinfo {author} {\bibfnamefont {D.}~\bibnamefont {McKay}}, \bibinfo
  {author} {\bibfnamefont {S.~Q.}\ \bibnamefont {Zhou}}, \bibinfo {author}
  {\bibfnamefont {D.}~\bibnamefont {Ceperley}}, \ and\ \bibinfo {author}
  {\bibfnamefont {B.}~\bibnamefont {DeMarco}},\ }\href {\doibase
  10.1103/physrevlett.102.055301} {\bibfield  {journal} {\bibinfo  {journal}
  {Physical Review Letters}\ }\textbf {\bibinfo {volume} {102}} (\bibinfo
  {year} {2009}),\ 10.1103/physrevlett.102.055301}\BibitemShut {NoStop}%
\bibitem [{\citenamefont {Micklitz}\ \emph {et~al.}(2014)\citenamefont
  {Micklitz}, \citenamefont {M{\"u}ller},\ and\ \citenamefont
  {Altland}}]{Micklitz2014}%
  \BibitemOpen
  \bibfield  {author} {\bibinfo {author} {\bibfnamefont {T.}~\bibnamefont
  {Micklitz}}, \bibinfo {author} {\bibfnamefont {C.~A.}\ \bibnamefont
  {M{\"u}ller}}, \ and\ \bibinfo {author} {\bibfnamefont {A.}~\bibnamefont
  {Altland}},\ }\href {\doibase 10.1103/physrevlett.112.110602} {\bibfield
  {journal} {\bibinfo  {journal} {Physical Review Letters}\ }\textbf {\bibinfo
  {volume} {112}} (\bibinfo {year} {2014}),\
  10.1103/physrevlett.112.110602}\BibitemShut {NoStop}%
\bibitem [{\citenamefont {Pons}\ and\ \citenamefont
  {Sanpera}(2017)}]{Pons2017}%
  \BibitemOpen
  \bibfield  {author} {\bibinfo {author} {\bibfnamefont {M.}~\bibnamefont
  {Pons}}\ and\ \bibinfo {author} {\bibfnamefont {A.}~\bibnamefont {Sanpera}},\
  }\href {\doibase 10.1103/physreva.95.033626} {\bibfield  {journal} {\bibinfo
  {journal} {Physical Review A}\ }\textbf {\bibinfo {volume} {95}} (\bibinfo
  {year} {2017}),\ 10.1103/physreva.95.033626}\BibitemShut {NoStop}%
\bibitem [{\citenamefont {Paiva}\ \emph {et~al.}(2015)\citenamefont {Paiva},
  \citenamefont {Khatami}, \citenamefont {Yang}, \citenamefont {Rousseau},
  \citenamefont {Jarrell}, \citenamefont {Moreno}, \citenamefont {Hulet},\ and\
  \citenamefont {Scalettar}}]{Paiva2015}%
  \BibitemOpen
  \bibfield  {author} {\bibinfo {author} {\bibfnamefont {T.}~\bibnamefont
  {Paiva}}, \bibinfo {author} {\bibfnamefont {E.}~\bibnamefont {Khatami}},
  \bibinfo {author} {\bibfnamefont {S.}~\bibnamefont {Yang}}, \bibinfo {author}
  {\bibfnamefont {V.}~\bibnamefont {Rousseau}}, \bibinfo {author}
  {\bibfnamefont {M.}~\bibnamefont {Jarrell}}, \bibinfo {author} {\bibfnamefont
  {J.}~\bibnamefont {Moreno}}, \bibinfo {author} {\bibfnamefont {R.~G.}\
  \bibnamefont {Hulet}}, \ and\ \bibinfo {author} {\bibfnamefont {R.~T.}\
  \bibnamefont {Scalettar}},\ }\href {\doibase 10.1103/physrevlett.115.240402}
  {\bibfield  {journal} {\bibinfo  {journal} {Physical Review Letters}\
  }\textbf {\bibinfo {volume} {115}} (\bibinfo {year} {2015}),\
  10.1103/physrevlett.115.240402}\BibitemShut {NoStop}%
\bibitem [{\citenamefont {Meldgin}\ \emph {et~al.}(2016)\citenamefont
  {Meldgin}, \citenamefont {Ray}, \citenamefont {Russ}, \citenamefont {Chen},
  \citenamefont {Ceperley},\ and\ \citenamefont {DeMarco}}]{Meldgin2016}%
  \BibitemOpen
  \bibfield  {author} {\bibinfo {author} {\bibfnamefont {C.}~\bibnamefont
  {Meldgin}}, \bibinfo {author} {\bibfnamefont {U.}~\bibnamefont {Ray}},
  \bibinfo {author} {\bibfnamefont {P.}~\bibnamefont {Russ}}, \bibinfo {author}
  {\bibfnamefont {D.}~\bibnamefont {Chen}}, \bibinfo {author} {\bibfnamefont
  {D.~M.}\ \bibnamefont {Ceperley}}, \ and\ \bibinfo {author} {\bibfnamefont
  {B.}~\bibnamefont {DeMarco}},\ }\href {\doibase 10.1038/nphys3695} {\bibfield
   {journal} {\bibinfo  {journal} {Nature Physics}\ }\textbf {\bibinfo {volume}
  {12}},\ \bibinfo {pages} {646} (\bibinfo {year} {2016})}\BibitemShut
  {NoStop}%
\bibitem [{\citenamefont {Volchkov}\ \emph {et~al.}(2018)\citenamefont
  {Volchkov}, \citenamefont {Pasek}, \citenamefont {Denechaud}, \citenamefont
  {Mukhtar}, \citenamefont {Aspect}, \citenamefont {Delande},\ and\
  \citenamefont {Josse}}]{Volchkov2018}%
  \BibitemOpen
  \bibfield  {author} {\bibinfo {author} {\bibfnamefont {V.~V.}\ \bibnamefont
  {Volchkov}}, \bibinfo {author} {\bibfnamefont {M.}~\bibnamefont {Pasek}},
  \bibinfo {author} {\bibfnamefont {V.}~\bibnamefont {Denechaud}}, \bibinfo
  {author} {\bibfnamefont {M.}~\bibnamefont {Mukhtar}}, \bibinfo {author}
  {\bibfnamefont {A.}~\bibnamefont {Aspect}}, \bibinfo {author} {\bibfnamefont
  {D.}~\bibnamefont {Delande}}, \ and\ \bibinfo {author} {\bibfnamefont
  {V.}~\bibnamefont {Josse}},\ }\href {\doibase 10.1103/physrevlett.120.060404}
  {\bibfield  {journal} {\bibinfo  {journal} {Physical Review Letters}\
  }\textbf {\bibinfo {volume} {120}} (\bibinfo {year} {2018}),\
  10.1103/physrevlett.120.060404}\BibitemShut {NoStop}%
\bibitem [{\citenamefont {Gersch}\ and\ \citenamefont
  {Knollman}(1963)}]{Gersch1963}%
  \BibitemOpen
  \bibfield  {author} {\bibinfo {author} {\bibfnamefont {H.~A.}\ \bibnamefont
  {Gersch}}\ and\ \bibinfo {author} {\bibfnamefont {G.~C.}\ \bibnamefont
  {Knollman}},\ }\href {\doibase 10.1103/physrev.129.959} {\bibfield  {journal}
  {\bibinfo  {journal} {Physical Review}\ }\textbf {\bibinfo {volume} {129}},\
  \bibinfo {pages} {959} (\bibinfo {year} {1963})}\BibitemShut {NoStop}%
\bibitem [{\citenamefont {Gavish}\ and\ \citenamefont
  {Castin}(2005)}]{Gavish2005}%
  \BibitemOpen
  \bibfield  {author} {\bibinfo {author} {\bibfnamefont {U.}~\bibnamefont
  {Gavish}}\ and\ \bibinfo {author} {\bibfnamefont {Y.}~\bibnamefont
  {Castin}},\ }\href {\doibase 10.1103/physrevlett.95.020401} {\bibfield
  {journal} {\bibinfo  {journal} {Physical Review Letters}\ }\textbf {\bibinfo
  {volume} {95}} (\bibinfo {year} {2005}),\
  10.1103/physrevlett.95.020401}\BibitemShut {NoStop}%
\bibitem [{\citenamefont {Calzetta}\ \emph {et~al.}(2006)\citenamefont
  {Calzetta}, \citenamefont {Hu},\ and\ \citenamefont {Rey}}]{Calzetta2006}%
  \BibitemOpen
  \bibfield  {author} {\bibinfo {author} {\bibfnamefont {E.}~\bibnamefont
  {Calzetta}}, \bibinfo {author} {\bibfnamefont {B.~L.}\ \bibnamefont {Hu}}, \
  and\ \bibinfo {author} {\bibfnamefont {A.~M.}\ \bibnamefont {Rey}},\ }\href
  {\doibase 10.1103/physreva.73.023610} {\bibfield  {journal} {\bibinfo
  {journal} {Physical Review A}\ }\textbf {\bibinfo {volume} {73}} (\bibinfo
  {year} {2006}),\ 10.1103/physreva.73.023610}\BibitemShut {NoStop}%
\bibitem [{\citenamefont {Tilahun}\ \emph {et~al.}(2011)\citenamefont
  {Tilahun}, \citenamefont {Duine},\ and\ \citenamefont
  {MacDonald}}]{Tilahun2011}%
  \BibitemOpen
  \bibfield  {author} {\bibinfo {author} {\bibfnamefont {D.}~\bibnamefont
  {Tilahun}}, \bibinfo {author} {\bibfnamefont {R.~A.}\ \bibnamefont {Duine}},
  \ and\ \bibinfo {author} {\bibfnamefont {A.~H.}\ \bibnamefont {MacDonald}},\
  }\href {\doibase 10.1103/physreva.84.033622} {\bibfield  {journal} {\bibinfo
  {journal} {Physical Review A}\ }\textbf {\bibinfo {volume} {84}} (\bibinfo
  {year} {2011}),\ 10.1103/physreva.84.033622}\BibitemShut {NoStop}%
\bibitem [{\citenamefont {Capogrosso-Sansone}\ \emph
  {et~al.}(2007)\citenamefont {Capogrosso-Sansone}, \citenamefont {Prokof'ev},\
  and\ \citenamefont {Svistunov}}]{Capogrosso-Sansone2007}%
  \BibitemOpen
  \bibfield  {author} {\bibinfo {author} {\bibfnamefont {B.}~\bibnamefont
  {Capogrosso-Sansone}}, \bibinfo {author} {\bibfnamefont {N.~V.}\ \bibnamefont
  {Prokof'ev}}, \ and\ \bibinfo {author} {\bibfnamefont {B.~V.}\ \bibnamefont
  {Svistunov}},\ }\href {\doibase 10.1103/physrevb.75.134302} {\bibfield
  {journal} {\bibinfo  {journal} {Physical Review B}\ }\textbf {\bibinfo
  {volume} {75}} (\bibinfo {year} {2007}),\
  10.1103/physrevb.75.134302}\BibitemShut {NoStop}%
\bibitem [{\citenamefont {{\L}{\k{a}}cki}\ \emph {et~al.}(2016)\citenamefont
  {{\L}{\k{a}}cki}, \citenamefont {Damski},\ and\ \citenamefont
  {Zakrzewski}}]{Lacki2016}%
  \BibitemOpen
  \bibfield  {author} {\bibinfo {author} {\bibfnamefont {M.}~\bibnamefont
  {{\L}{\k{a}}cki}}, \bibinfo {author} {\bibfnamefont {B.}~\bibnamefont
  {Damski}}, \ and\ \bibinfo {author} {\bibfnamefont {J.}~\bibnamefont
  {Zakrzewski}},\ }\href {\doibase 10.1038/srep38340} {\bibfield  {journal}
  {\bibinfo  {journal} {Scientific Reports}\ }\textbf {\bibinfo {volume} {6}}
  (\bibinfo {year} {2016}),\ 10.1038/srep38340}\BibitemShut {NoStop}%
\bibitem [{\citenamefont {Lewenstein}\ \emph {et~al.}(2007)\citenamefont
  {Lewenstein}, \citenamefont {Sanpera}, \citenamefont {Ahufinger},
  \citenamefont {Damski}, \citenamefont {Sen(De)},\ and\ \citenamefont
  {Sen}}]{Lewenstein2007}%
  \BibitemOpen
  \bibfield  {author} {\bibinfo {author} {\bibfnamefont {M.}~\bibnamefont
  {Lewenstein}}, \bibinfo {author} {\bibfnamefont {A.}~\bibnamefont {Sanpera}},
  \bibinfo {author} {\bibfnamefont {V.}~\bibnamefont {Ahufinger}}, \bibinfo
  {author} {\bibfnamefont {B.}~\bibnamefont {Damski}}, \bibinfo {author}
  {\bibfnamefont {A.}~\bibnamefont {Sen(De)}}, \ and\ \bibinfo {author}
  {\bibfnamefont {U.}~\bibnamefont {Sen}},\ }\href {\doibase
  10.1080/00018730701223200} {\bibfield  {journal} {\bibinfo  {journal}
  {Advances in Physics}\ }\textbf {\bibinfo {volume} {56}},\ \bibinfo {pages}
  {243} (\bibinfo {year} {2007})},\ \Eprint
  {http://arxiv.org/abs/https://doi.org/10.1080/00018730701223200}
  {https://doi.org/10.1080/00018730701223200} \BibitemShut {NoStop}%
\bibitem [{\citenamefont {Orzel}\ \emph {et~al.}(2001)\citenamefont {Orzel},
  \citenamefont {Tuchman}, \citenamefont {Fenselau}, \citenamefont {Yasuda},\
  and\ \citenamefont {Kasevich}}]{Orzel2001}%
  \BibitemOpen
  \bibfield  {author} {\bibinfo {author} {\bibfnamefont {C.}~\bibnamefont
  {Orzel}}, \bibinfo {author} {\bibfnamefont {A.~K.}\ \bibnamefont {Tuchman}},
  \bibinfo {author} {\bibfnamefont {M.~L.}\ \bibnamefont {Fenselau}}, \bibinfo
  {author} {\bibfnamefont {M.}~\bibnamefont {Yasuda}}, \ and\ \bibinfo {author}
  {\bibfnamefont {M.~A.}\ \bibnamefont {Kasevich}},\ }\href {\doibase
  10.1126/science.1058149} {\bibfield  {journal} {\bibinfo  {journal}
  {Science}\ }\textbf {\bibinfo {volume} {291}},\ \bibinfo {pages} {2386}
  (\bibinfo {year} {2001})}\BibitemShut {NoStop}%
\bibitem [{\citenamefont {Bloch}\ \emph {et~al.}(2008)\citenamefont {Bloch},
  \citenamefont {Dalibard},\ and\ \citenamefont {Zwerger}}]{Bloch2008}%
  \BibitemOpen
  \bibfield  {author} {\bibinfo {author} {\bibfnamefont {I.}~\bibnamefont
  {Bloch}}, \bibinfo {author} {\bibfnamefont {J.}~\bibnamefont {Dalibard}}, \
  and\ \bibinfo {author} {\bibfnamefont {W.}~\bibnamefont {Zwerger}},\ }\href
  {\doibase 10.1103/revmodphys.80.885} {\bibfield  {journal} {\bibinfo
  {journal} {Reviews of Modern Physics}\ }\textbf {\bibinfo {volume} {80}},\
  \bibinfo {pages} {885} (\bibinfo {year} {2008})}\BibitemShut {NoStop}%
\bibitem [{\citenamefont {Zhang}\ \emph {et~al.}(2012)\citenamefont {Zhang},
  \citenamefont {Chen}, \citenamefont {Liu}, \citenamefont {Wu}, \citenamefont
  {Wen},\ and\ \citenamefont {Liu}}]{Zhang2012}%
  \BibitemOpen
  \bibfield  {author} {\bibinfo {author} {\bibfnamefont {X.}~\bibnamefont
  {Zhang}}, \bibinfo {author} {\bibfnamefont {Y.}~\bibnamefont {Chen}},
  \bibinfo {author} {\bibfnamefont {G.}~\bibnamefont {Liu}}, \bibinfo {author}
  {\bibfnamefont {W.}~\bibnamefont {Wu}}, \bibinfo {author} {\bibfnamefont
  {L.}~\bibnamefont {Wen}}, \ and\ \bibinfo {author} {\bibfnamefont
  {W.}~\bibnamefont {Liu}},\ }\href {\doibase 10.1007/s11434-012-5095-1}
  {\bibfield  {journal} {\bibinfo  {journal} {Chinese Science Bulletin}\
  }\textbf {\bibinfo {volume} {57}},\ \bibinfo {pages} {1910} (\bibinfo {year}
  {2012})}\BibitemShut {NoStop}%
\bibitem [{\citenamefont {Ohmori}\ \emph {et~al.}(2017)\citenamefont {Ohmori},
  \citenamefont {Pupillo}, \citenamefont {Thywissen},\ and\ \citenamefont
  {Weidem{\"u}ller}}]{Ohmori2017}%
  \BibitemOpen
  \bibfield  {author} {\bibinfo {author} {\bibfnamefont {K.}~\bibnamefont
  {Ohmori}}, \bibinfo {author} {\bibfnamefont {G.}~\bibnamefont {Pupillo}},
  \bibinfo {author} {\bibfnamefont {J.~H.}\ \bibnamefont {Thywissen}}, \ and\
  \bibinfo {author} {\bibfnamefont {M.}~\bibnamefont {Weidem{\"u}ller}},\
  }\href {\doibase 10.1088/1361-6455/aa9d13} {\bibfield  {journal} {\bibinfo
  {journal} {Journal of Physics B: Atomic, Molecular and Optical Physics}\
  }\textbf {\bibinfo {volume} {51}},\ \bibinfo {pages} {020201} (\bibinfo
  {year} {2017})}\BibitemShut {NoStop}%
\bibitem [{\citenamefont {Hofstetter}(2006)}]{Hofstetter2006}%
  \BibitemOpen
  \bibfield  {author} {\bibinfo {author} {\bibfnamefont {W.}~\bibnamefont
  {Hofstetter}},\ }\href {\doibase 10.1080/14786430500228770} {\bibfield
  {journal} {\bibinfo  {journal} {Philosophical Magazine}\ }\textbf {\bibinfo
  {volume} {86}},\ \bibinfo {pages} {1891} (\bibinfo {year}
  {2006})}\BibitemShut {NoStop}%
\bibitem [{\citenamefont {Esslinger}(2010)}]{Esslinger2010}%
  \BibitemOpen
  \bibfield  {author} {\bibinfo {author} {\bibfnamefont {T.}~\bibnamefont
  {Esslinger}},\ }\href {\doibase 10.1146/annurev-conmatphys-070909-104059}
  {\bibfield  {journal} {\bibinfo  {journal} {Annual Review of Condensed Matter
  Physics}\ }\textbf {\bibinfo {volume} {1}},\ \bibinfo {pages} {129} (\bibinfo
  {year} {2010})}\BibitemShut {NoStop}%
\bibitem [{\citenamefont {Greiner}\ \emph {et~al.}(2002)\citenamefont
  {Greiner}, \citenamefont {Mandel}, \citenamefont {Esslinger}, \citenamefont
  {Hansch},\ and\ \citenamefont {Bloch}}]{Greiner2002}%
  \BibitemOpen
  \bibfield  {author} {\bibinfo {author} {\bibfnamefont {M.}~\bibnamefont
  {Greiner}}, \bibinfo {author} {\bibfnamefont {O.}~\bibnamefont {Mandel}},
  \bibinfo {author} {\bibfnamefont {T.}~\bibnamefont {Esslinger}}, \bibinfo
  {author} {\bibfnamefont {T.~W.}\ \bibnamefont {Hansch}}, \ and\ \bibinfo
  {author} {\bibfnamefont {I.}~\bibnamefont {Bloch}},\ }\href {\doibase
  10.1038/415039a} {\bibfield  {journal} {\bibinfo  {journal} {Nature}\
  }\textbf {\bibinfo {volume} {415}},\ \bibinfo {pages} {39} (\bibinfo {year}
  {2002})}\BibitemShut {NoStop}%
\bibitem [{\citenamefont {Chen}\ \emph {et~al.}(2011)\citenamefont {Chen},
  \citenamefont {White}, \citenamefont {Borries},\ and\ \citenamefont
  {DeMarco}}]{Chen2011}%
  \BibitemOpen
  \bibfield  {author} {\bibinfo {author} {\bibfnamefont {D.}~\bibnamefont
  {Chen}}, \bibinfo {author} {\bibfnamefont {M.}~\bibnamefont {White}},
  \bibinfo {author} {\bibfnamefont {C.}~\bibnamefont {Borries}}, \ and\
  \bibinfo {author} {\bibfnamefont {B.}~\bibnamefont {DeMarco}},\ }\href
  {\doibase 10.1103/physrevlett.106.235304} {\bibfield  {journal} {\bibinfo
  {journal} {Physical Review Letters}\ }\textbf {\bibinfo {volume} {106}}
  (\bibinfo {year} {2011}),\ 10.1103/physrevlett.106.235304}\BibitemShut
  {NoStop}%
\bibitem [{\citenamefont {Trotzky}\ \emph {et~al.}(2010)\citenamefont
  {Trotzky}, \citenamefont {Pollet}, \citenamefont {Gerbier}, \citenamefont
  {Schnorrberger}, \citenamefont {Bloch}, \citenamefont {Prokof'ev},
  \citenamefont {Svistunov},\ and\ \citenamefont {Troyer}}]{Trotzky2010}%
  \BibitemOpen
  \bibfield  {author} {\bibinfo {author} {\bibfnamefont {S.}~\bibnamefont
  {Trotzky}}, \bibinfo {author} {\bibfnamefont {L.}~\bibnamefont {Pollet}},
  \bibinfo {author} {\bibfnamefont {F.}~\bibnamefont {Gerbier}}, \bibinfo
  {author} {\bibfnamefont {U.}~\bibnamefont {Schnorrberger}}, \bibinfo {author}
  {\bibfnamefont {I.}~\bibnamefont {Bloch}}, \bibinfo {author} {\bibfnamefont
  {N.~V.}\ \bibnamefont {Prokof'ev}}, \bibinfo {author} {\bibfnamefont
  {B.}~\bibnamefont {Svistunov}}, \ and\ \bibinfo {author} {\bibfnamefont
  {M.}~\bibnamefont {Troyer}},\ }\href {\doibase 10.1038/nphys1799} {\bibfield
  {journal} {\bibinfo  {journal} {Nature Physics}\ }\textbf {\bibinfo {volume}
  {6}},\ \bibinfo {pages} {998} (\bibinfo {year} {2010})}\BibitemShut {NoStop}%
\bibitem [{\citenamefont {McKay}\ \emph {et~al.}(2015)\citenamefont {McKay},
  \citenamefont {Ray}, \citenamefont {Natu}, \citenamefont {Russ},
  \citenamefont {Ceperley},\ and\ \citenamefont {DeMarco}}]{McKay2015}%
  \BibitemOpen
  \bibfield  {author} {\bibinfo {author} {\bibfnamefont {D.}~\bibnamefont
  {McKay}}, \bibinfo {author} {\bibfnamefont {U.}~\bibnamefont {Ray}}, \bibinfo
  {author} {\bibfnamefont {S.}~\bibnamefont {Natu}}, \bibinfo {author}
  {\bibfnamefont {P.}~\bibnamefont {Russ}}, \bibinfo {author} {\bibfnamefont
  {D.}~\bibnamefont {Ceperley}}, \ and\ \bibinfo {author} {\bibfnamefont
  {B.}~\bibnamefont {DeMarco}},\ }\href {\doibase 10.1103/physreva.91.023625}
  {\bibfield  {journal} {\bibinfo  {journal} {Physical Review A}\ }\textbf
  {\bibinfo {volume} {91}} (\bibinfo {year} {2015}),\
  10.1103/physreva.91.023625}\BibitemShut {NoStop}%
\bibitem [{\citenamefont {Zhou}\ and\ \citenamefont
  {Ceperley}(2010)}]{Zhou2010}%
  \BibitemOpen
  \bibfield  {author} {\bibinfo {author} {\bibfnamefont {S.~Q.}\ \bibnamefont
  {Zhou}}\ and\ \bibinfo {author} {\bibfnamefont {D.~M.}\ \bibnamefont
  {Ceperley}},\ }\href {\doibase 10.1103/physreva.81.013402} {\bibfield
  {journal} {\bibinfo  {journal} {Physical Review A}\ }\textbf {\bibinfo
  {volume} {81}} (\bibinfo {year} {2010}),\
  10.1103/physreva.81.013402}\BibitemShut {NoStop}%
\bibitem [{\citenamefont {Ray}(2015)}]{RayDissertation:2015}%
  \BibitemOpen
  \bibfield  {author} {\bibinfo {author} {\bibfnamefont {U.}~\bibnamefont
  {Ray}},\ }\emph {\bibinfo {title} {Properties of dirty bosons in disordered
  optical lattices}},\ \href {http://hdl.handle.net/2142/78442} {Ph.D.
  thesis},\ \bibinfo  {school} {University of Illinois at Urbana-Champaign}
  (\bibinfo {year} {2015})\BibitemShut {NoStop}%
\bibitem [{\citenamefont {Harris}\ and\ \citenamefont
  {Lubensky}(1974)}]{Harris1974}%
  \BibitemOpen
  \bibfield  {author} {\bibinfo {author} {\bibfnamefont {A.~B.}\ \bibnamefont
  {Harris}}\ and\ \bibinfo {author} {\bibfnamefont {T.~C.}\ \bibnamefont
  {Lubensky}},\ }\href {\doibase 10.1103/physrevlett.33.1540} {\bibfield
  {journal} {\bibinfo  {journal} {Physical Review Letters}\ }\textbf {\bibinfo
  {volume} {33}},\ \bibinfo {pages} {1540} (\bibinfo {year}
  {1974})}\BibitemShut {NoStop}%
\bibitem [{\citenamefont {Chayes}\ \emph {et~al.}(1986)\citenamefont {Chayes},
  \citenamefont {Chayes}, \citenamefont {Fisher},\ and\ \citenamefont
  {Spencer}}]{Chayes1986}%
  \BibitemOpen
  \bibfield  {author} {\bibinfo {author} {\bibfnamefont {J.~T.}\ \bibnamefont
  {Chayes}}, \bibinfo {author} {\bibfnamefont {L.}~\bibnamefont {Chayes}},
  \bibinfo {author} {\bibfnamefont {D.~S.}\ \bibnamefont {Fisher}}, \ and\
  \bibinfo {author} {\bibfnamefont {T.}~\bibnamefont {Spencer}},\ }\href
  {\doibase 10.1103/physrevlett.57.2999} {\bibfield  {journal} {\bibinfo
  {journal} {Physical Review Letters}\ }\textbf {\bibinfo {volume} {57}},\
  \bibinfo {pages} {2999} (\bibinfo {year} {1986})}\BibitemShut {NoStop}%
\bibitem [{\citenamefont {Aharony}\ and\ \citenamefont
  {Harris}(1996)}]{Aharony1996}%
  \BibitemOpen
  \bibfield  {author} {\bibinfo {author} {\bibfnamefont {A.}~\bibnamefont
  {Aharony}}\ and\ \bibinfo {author} {\bibfnamefont {A.~B.}\ \bibnamefont
  {Harris}},\ }\href {\doibase 10.1103/physrevlett.77.3700} {\bibfield
  {journal} {\bibinfo  {journal} {Physical Review Letters}\ }\textbf {\bibinfo
  {volume} {77}},\ \bibinfo {pages} {3700} (\bibinfo {year}
  {1996})}\BibitemShut {NoStop}%
\bibitem [{\citenamefont {Wiseman}\ and\ \citenamefont
  {Domany}(1998)}]{Wiseman1998}%
  \BibitemOpen
  \bibfield  {author} {\bibinfo {author} {\bibfnamefont {S.}~\bibnamefont
  {Wiseman}}\ and\ \bibinfo {author} {\bibfnamefont {E.}~\bibnamefont
  {Domany}},\ }\href {\doibase 10.1103/physrevlett.81.22} {\bibfield  {journal}
  {\bibinfo  {journal} {Physical Review Letters}\ }\textbf {\bibinfo {volume}
  {81}},\ \bibinfo {pages} {22} (\bibinfo {year} {1998})}\BibitemShut {NoStop}%
\bibitem [{\citenamefont {Harris}(1974)}]{Harris1974a}%
  \BibitemOpen
  \bibfield  {author} {\bibinfo {author} {\bibfnamefont {A.~B.}\ \bibnamefont
  {Harris}},\ }\href {\doibase 10.1088/0022-3719/7/9/009} {\bibfield  {journal}
  {\bibinfo  {journal} {Journal of Physics C: Solid State Physics}\ }\textbf
  {\bibinfo {volume} {7}},\ \bibinfo {pages} {1671} (\bibinfo {year}
  {1974})}\BibitemShut {NoStop}%
\bibitem [{\citenamefont {Vojta}(2006)}]{Vojta2006}%
  \BibitemOpen
  \bibfield  {author} {\bibinfo {author} {\bibfnamefont {T.}~\bibnamefont
  {Vojta}},\ }\href {\doibase 10.1088/0305-4470/39/22/r01} {\bibfield
  {journal} {\bibinfo  {journal} {Journal of Physics A: Mathematical and
  General}\ }\textbf {\bibinfo {volume} {39}},\ \bibinfo {pages} {R143}
  (\bibinfo {year} {2006})}\BibitemShut {NoStop}%
\bibitem [{\citenamefont {Kisker}\ and\ \citenamefont
  {Rieger}(1997)}]{Kisker1997}%
  \BibitemOpen
  \bibfield  {author} {\bibinfo {author} {\bibfnamefont {J.}~\bibnamefont
  {Kisker}}\ and\ \bibinfo {author} {\bibfnamefont {H.}~\bibnamefont
  {Rieger}},\ }\href {\doibase 10.1103/physrevb.55.r11981} {\bibfield
  {journal} {\bibinfo  {journal} {Physical Review B}\ }\textbf {\bibinfo
  {volume} {55}},\ \bibinfo {pages} {R11981} (\bibinfo {year}
  {1997})}\BibitemShut {NoStop}%
\bibitem [{\citenamefont {Kr{\"u}ger}\ \emph {et~al.}(2011)\citenamefont
  {Kr{\"u}ger}, \citenamefont {Hong},\ and\ \citenamefont
  {Phillips}}]{Krueger2011}%
  \BibitemOpen
  \bibfield  {author} {\bibinfo {author} {\bibfnamefont {F.}~\bibnamefont
  {Kr{\"u}ger}}, \bibinfo {author} {\bibfnamefont {S.}~\bibnamefont {Hong}}, \
  and\ \bibinfo {author} {\bibfnamefont {P.}~\bibnamefont {Phillips}},\ }\href
  {\doibase 10.1103/physrevb.84.115118} {\bibfield  {journal} {\bibinfo
  {journal} {Physical Review B}\ }\textbf {\bibinfo {volume} {84}} (\bibinfo
  {year} {2011}),\ 10.1103/physrevb.84.115118}\BibitemShut {NoStop}%
\bibitem [{\citenamefont {Yao}\ \emph {et~al.}(2014)\citenamefont {Yao},
  \citenamefont {da~Costa}, \citenamefont {Kiselev},\ and\ \citenamefont
  {Prokof'ev}}]{Yao2014}%
  \BibitemOpen
  \bibfield  {author} {\bibinfo {author} {\bibfnamefont {Z.}~\bibnamefont
  {Yao}}, \bibinfo {author} {\bibfnamefont {K.~P.~C.}\ \bibnamefont
  {da~Costa}}, \bibinfo {author} {\bibfnamefont {M.}~\bibnamefont {Kiselev}}, \
  and\ \bibinfo {author} {\bibfnamefont {N.}~\bibnamefont {Prokof'ev}},\ }\href
  {\doibase 10.1103/physrevlett.112.225301} {\bibfield  {journal} {\bibinfo
  {journal} {Physical Review Letters}\ }\textbf {\bibinfo {volume} {112}}
  (\bibinfo {year} {2014}),\ 10.1103/physrevlett.112.225301}\BibitemShut
  {NoStop}%
\bibitem [{\citenamefont {Bissbort}\ \emph {et~al.}(2010)\citenamefont
  {Bissbort}, \citenamefont {Thomale},\ and\ \citenamefont
  {Hofstetter}}]{Bissbort2010}%
  \BibitemOpen
  \bibfield  {author} {\bibinfo {author} {\bibfnamefont {U.}~\bibnamefont
  {Bissbort}}, \bibinfo {author} {\bibfnamefont {R.}~\bibnamefont {Thomale}}, \
  and\ \bibinfo {author} {\bibfnamefont {W.}~\bibnamefont {Hofstetter}},\
  }\href {\doibase 10.1103/physreva.81.063643} {\bibfield  {journal} {\bibinfo
  {journal} {Physical Review A}\ }\textbf {\bibinfo {volume} {81}} (\bibinfo
  {year} {2010}),\ 10.1103/physreva.81.063643}\BibitemShut {NoStop}%
\bibitem [{\citenamefont {Stauffer}\ and\ \citenamefont
  {Aharony}(1994)}]{Stauffer:1994}%
  \BibitemOpen
  \bibfield  {author} {\bibinfo {author} {\bibfnamefont {D.}~\bibnamefont
  {Stauffer}}\ and\ \bibinfo {author} {\bibfnamefont {A.}~\bibnamefont
  {Aharony}},\ }\href {https://books.google.com/books?id=v66plleij5QC} {\emph
  {\bibinfo {title} {{Introduction To Percolation Theory}}}}\ (\bibinfo
  {publisher} {Taylor \& Francis},\ \bibinfo {year} {1994})\BibitemShut
  {NoStop}%
\bibitem [{\citenamefont {Leggett}(2006)}]{LeggettBook2006}%
  \BibitemOpen
  \bibfield  {author} {\bibinfo {author} {\bibfnamefont {A.~J.}\ \bibnamefont
  {Leggett}},\ }\href@noop {} {\emph {\bibinfo {title} {Quantum Liquids: Bose
  Condensation and Cooper Pairing in Condensed-Matter Systems (Oxford Graduate
  Texts)}}}\ (\bibinfo  {publisher} {Oxford University Press},\ \bibinfo {year}
  {2006})\BibitemShut {NoStop}%
\bibitem [{\citenamefont {Weinrib}\ and\ \citenamefont
  {Halperin}(1983)}]{Weinrib1983}%
  \BibitemOpen
  \bibfield  {author} {\bibinfo {author} {\bibfnamefont {A.}~\bibnamefont
  {Weinrib}}\ and\ \bibinfo {author} {\bibfnamefont {B.~I.}\ \bibnamefont
  {Halperin}},\ }\href {\doibase 10.1103/physrevb.27.413} {\bibfield  {journal}
  {\bibinfo  {journal} {Physical Review B}\ }\textbf {\bibinfo {volume} {27}},\
  \bibinfo {pages} {413} (\bibinfo {year} {1983})}\BibitemShut {NoStop}%
\bibitem [{\citenamefont {Granato}\ and\ \citenamefont
  {Kosterlitz}(1986)}]{Granato1986}%
  \BibitemOpen
  \bibfield  {author} {\bibinfo {author} {\bibfnamefont {E.}~\bibnamefont
  {Granato}}\ and\ \bibinfo {author} {\bibfnamefont {J.~M.}\ \bibnamefont
  {Kosterlitz}},\ }\href {\doibase 10.1103/physrevb.33.6533} {\bibfield
  {journal} {\bibinfo  {journal} {Physical Review B}\ }\textbf {\bibinfo
  {volume} {33}},\ \bibinfo {pages} {6533} (\bibinfo {year}
  {1986})}\BibitemShut {NoStop}%
\bibitem [{\citenamefont {Fisher}\ \emph {et~al.}(1991)\citenamefont {Fisher},
  \citenamefont {Fisher},\ and\ \citenamefont {Huse}}]{Fisher1991}%
  \BibitemOpen
  \bibfield  {author} {\bibinfo {author} {\bibfnamefont {D.~S.}\ \bibnamefont
  {Fisher}}, \bibinfo {author} {\bibfnamefont {M.~P.~A.}\ \bibnamefont
  {Fisher}}, \ and\ \bibinfo {author} {\bibfnamefont {D.~A.}\ \bibnamefont
  {Huse}},\ }\href {\doibase 10.1103/physrevb.43.130} {\bibfield  {journal}
  {\bibinfo  {journal} {Physical Review B}\ }\textbf {\bibinfo {volume} {43}},\
  \bibinfo {pages} {130} (\bibinfo {year} {1991})}\BibitemShut {NoStop}%
\bibitem [{\citenamefont {Fisher}\ \emph {et~al.}(2004)\citenamefont {Fisher},
  \citenamefont {Kalinov}, \citenamefont {Voloshin}, \citenamefont
  {Babushkina}, \citenamefont {Khomskii}, \citenamefont {Zhang},\ and\
  \citenamefont {Palstra}}]{Fisher2004}%
  \BibitemOpen
  \bibfield  {author} {\bibinfo {author} {\bibfnamefont {L.~M.}\ \bibnamefont
  {Fisher}}, \bibinfo {author} {\bibfnamefont {A.~V.}\ \bibnamefont {Kalinov}},
  \bibinfo {author} {\bibfnamefont {I.~F.}\ \bibnamefont {Voloshin}}, \bibinfo
  {author} {\bibfnamefont {N.~A.}\ \bibnamefont {Babushkina}}, \bibinfo
  {author} {\bibfnamefont {D.~I.}\ \bibnamefont {Khomskii}}, \bibinfo {author}
  {\bibfnamefont {Y.}~\bibnamefont {Zhang}}, \ and\ \bibinfo {author}
  {\bibfnamefont {T.~T.~M.}\ \bibnamefont {Palstra}},\ }\href {\doibase
  10.1103/physrevb.70.212411} {\bibfield  {journal} {\bibinfo  {journal}
  {Physical Review B}\ }\textbf {\bibinfo {volume} {70}} (\bibinfo {year}
  {2004}),\ 10.1103/physrevb.70.212411}\BibitemShut {NoStop}%
\bibitem [{\citenamefont {Maestro}\ \emph {et~al.}(2006)\citenamefont
  {Maestro}, \citenamefont {Rosenow},\ and\ \citenamefont
  {Sachdev}}]{Maestro2006}%
  \BibitemOpen
  \bibfield  {author} {\bibinfo {author} {\bibfnamefont {A.~D.}\ \bibnamefont
  {Maestro}}, \bibinfo {author} {\bibfnamefont {B.}~\bibnamefont {Rosenow}}, \
  and\ \bibinfo {author} {\bibfnamefont {S.}~\bibnamefont {Sachdev}},\ }\href
  {\doibase 10.1103/physrevb.74.024520} {\bibfield  {journal} {\bibinfo
  {journal} {Physical Review B}\ }\textbf {\bibinfo {volume} {74}} (\bibinfo
  {year} {2006}),\ 10.1103/physrevb.74.024520}\BibitemShut {NoStop}%
\bibitem [{\citenamefont {Pasienski}\ \emph {et~al.}(2010)\citenamefont
  {Pasienski}, \citenamefont {McKay}, \citenamefont {White},\ and\
  \citenamefont {DeMarco}}]{Pasienski2010}%
  \BibitemOpen
  \bibfield  {author} {\bibinfo {author} {\bibfnamefont {M.}~\bibnamefont
  {Pasienski}}, \bibinfo {author} {\bibfnamefont {D.}~\bibnamefont {McKay}},
  \bibinfo {author} {\bibfnamefont {M.}~\bibnamefont {White}}, \ and\ \bibinfo
  {author} {\bibfnamefont {B.}~\bibnamefont {DeMarco}},\ }\href {\doibase
  10.1038/nphys1726} {\bibfield  {journal} {\bibinfo  {journal} {Nature
  Physics}\ }\textbf {\bibinfo {volume} {6}},\ \bibinfo {pages} {677} (\bibinfo
  {year} {2010})}\BibitemShut {NoStop}%
\bibitem [{\citenamefont {Vojta}(2013)}]{Vojta2013}%
  \BibitemOpen
  \bibfield  {author} {\bibinfo {author} {\bibfnamefont {T.}~\bibnamefont
  {Vojta}},\ }\href {\doibase 10.1063/1.4818403} {\bibfield  {journal}
  {\bibinfo  {journal} {AIP Conference Proceedings}\ }\textbf {\bibinfo
  {volume} {1550}},\ \bibinfo {pages} {188} (\bibinfo {year} {2013})},\ \Eprint
  {http://arxiv.org/abs/https://aip.scitation.org/doi/pdf/10.1063/1.4818403}
  {https://aip.scitation.org/doi/pdf/10.1063/1.4818403} \BibitemShut {NoStop}%
\bibitem [{\citenamefont {Wiseman}\ and\ \citenamefont
  {Domany}(1995)}]{Wiseman1995}%
  \BibitemOpen
  \bibfield  {author} {\bibinfo {author} {\bibfnamefont {S.}~\bibnamefont
  {Wiseman}}\ and\ \bibinfo {author} {\bibfnamefont {E.}~\bibnamefont
  {Domany}},\ }\href {\doibase 10.1103/physreve.52.3469} {\bibfield  {journal}
  {\bibinfo  {journal} {Physical Review E}\ }\textbf {\bibinfo {volume} {52}},\
  \bibinfo {pages} {3469} (\bibinfo {year} {1995})}\BibitemShut {NoStop}%
\bibitem [{\citenamefont {Martino}\ and\ \citenamefont
  {Giansanti}(1998)}]{Martino1998}%
  \BibitemOpen
  \bibfield  {author} {\bibinfo {author} {\bibfnamefont {A.~D.}\ \bibnamefont
  {Martino}}\ and\ \bibinfo {author} {\bibfnamefont {A.}~\bibnamefont
  {Giansanti}},\ }\href {\doibase 10.1088/0305-4470/31/44/005} {\bibfield
  {journal} {\bibinfo  {journal} {Journal of Physics A: Mathematical and
  General}\ }\textbf {\bibinfo {volume} {31}},\ \bibinfo {pages} {8757}
  (\bibinfo {year} {1998})}\BibitemShut {NoStop}%
\bibitem [{\citenamefont {Roy}\ and\ \citenamefont
  {Bhattacharjee}(2006)}]{Roy2006}%
  \BibitemOpen
  \bibfield  {author} {\bibinfo {author} {\bibfnamefont {S.}~\bibnamefont
  {Roy}}\ and\ \bibinfo {author} {\bibfnamefont {S.~M.}\ \bibnamefont
  {Bhattacharjee}},\ }\href {\doibase 10.1016/j.physleta.2005.10.105}
  {\bibfield  {journal} {\bibinfo  {journal} {Physics Letters A}\ }\textbf
  {\bibinfo {volume} {352}},\ \bibinfo {pages} {13} (\bibinfo {year}
  {2006})}\BibitemShut {NoStop}%
\bibitem [{\citenamefont {Hegg}\ \emph {et~al.}(2013)\citenamefont {Hegg},
  \citenamefont {Kruger},\ and\ \citenamefont {Phillips}}]{Hegg2013}%
  \BibitemOpen
  \bibfield  {author} {\bibinfo {author} {\bibfnamefont {A.}~\bibnamefont
  {Hegg}}, \bibinfo {author} {\bibfnamefont {F.}~\bibnamefont {Kruger}}, \ and\
  \bibinfo {author} {\bibfnamefont {P.~W.}\ \bibnamefont {Phillips}},\ }\href
  {\doibase 10.1103/physrevb.88.134206} {\bibfield  {journal} {\bibinfo
  {journal} {Physical Review B}\ }\textbf {\bibinfo {volume} {88}} (\bibinfo
  {year} {2013}),\ 10.1103/physrevb.88.134206}\BibitemShut {NoStop}%
\bibitem [{\citenamefont {Dotsenko}\ \emph {et~al.}(2017)\citenamefont
  {Dotsenko}, \citenamefont {Holovatch}, \citenamefont {Dudka},\ and\
  \citenamefont {Weigel}}]{Dotsenko2017}%
  \BibitemOpen
  \bibfield  {author} {\bibinfo {author} {\bibfnamefont {V.}~\bibnamefont
  {Dotsenko}}, \bibinfo {author} {\bibfnamefont {Y.}~\bibnamefont {Holovatch}},
  \bibinfo {author} {\bibfnamefont {M.}~\bibnamefont {Dudka}}, \ and\ \bibinfo
  {author} {\bibfnamefont {M.}~\bibnamefont {Weigel}},\ }\href {\doibase
  10.1103/physreve.95.032118} {\bibfield  {journal} {\bibinfo  {journal}
  {Physical Review E}\ }\textbf {\bibinfo {volume} {95}} (\bibinfo {year}
  {2017}),\ 10.1103/physreve.95.032118}\BibitemShut {NoStop}%
\bibitem [{\citenamefont {Fallani}\ \emph {et~al.}(2007)\citenamefont
  {Fallani}, \citenamefont {Lye}, \citenamefont {Guarrera}, \citenamefont
  {Fort},\ and\ \citenamefont {Inguscio}}]{Fallani2007}%
  \BibitemOpen
  \bibfield  {author} {\bibinfo {author} {\bibfnamefont {L.}~\bibnamefont
  {Fallani}}, \bibinfo {author} {\bibfnamefont {J.~E.}\ \bibnamefont {Lye}},
  \bibinfo {author} {\bibfnamefont {V.}~\bibnamefont {Guarrera}}, \bibinfo
  {author} {\bibfnamefont {C.}~\bibnamefont {Fort}}, \ and\ \bibinfo {author}
  {\bibfnamefont {M.}~\bibnamefont {Inguscio}},\ }\href {\doibase
  10.1103/physrevlett.98.130404} {\bibfield  {journal} {\bibinfo  {journal}
  {Physical Review Letters}\ }\textbf {\bibinfo {volume} {98}} (\bibinfo {year}
  {2007}),\ 10.1103/physrevlett.98.130404}\BibitemShut {NoStop}%
\bibitem [{\citenamefont {Sandvik}\ and\ \citenamefont
  {Kurkij{\"a}rvi}(1991)}]{Sandvik1991}%
  \BibitemOpen
  \bibfield  {author} {\bibinfo {author} {\bibfnamefont {A.~W.}\ \bibnamefont
  {Sandvik}}\ and\ \bibinfo {author} {\bibfnamefont {J.}~\bibnamefont
  {Kurkij{\"a}rvi}},\ }\href {\doibase 10.1103/physrevb.43.5950} {\bibfield
  {journal} {\bibinfo  {journal} {Physical Review B}\ }\textbf {\bibinfo
  {volume} {43}},\ \bibinfo {pages} {5950} (\bibinfo {year}
  {1991})}\BibitemShut {NoStop}%
\bibitem [{\citenamefont {Sandvik}(1992)}]{Sandvik1992}%
  \BibitemOpen
  \bibfield  {author} {\bibinfo {author} {\bibfnamefont {A.~W.}\ \bibnamefont
  {Sandvik}},\ }\href {\doibase 10.1088/0305-4470/25/13/017} {\bibfield
  {journal} {\bibinfo  {journal} {Journal of Physics A: Mathematical and
  General}\ }\textbf {\bibinfo {volume} {25}},\ \bibinfo {pages} {3667}
  (\bibinfo {year} {1992})}\BibitemShut {NoStop}%
\bibitem [{\citenamefont {Sandvik}(1999)}]{Sandvik1999}%
  \BibitemOpen
  \bibfield  {author} {\bibinfo {author} {\bibfnamefont {A.~W.}\ \bibnamefont
  {Sandvik}},\ }\href {\doibase 10.1103/physrevb.59.r14157} {\bibfield
  {journal} {\bibinfo  {journal} {Physical Review B}\ }\textbf {\bibinfo
  {volume} {59}},\ \bibinfo {pages} {R14157} (\bibinfo {year}
  {1999})}\BibitemShut {NoStop}%
\bibitem [{\citenamefont {Syljuasen}\ and\ \citenamefont
  {Sandvik}(2002)}]{Syljuaasen2002}%
  \BibitemOpen
  \bibfield  {author} {\bibinfo {author} {\bibfnamefont {O.~F.}\ \bibnamefont
  {Syljuasen}}\ and\ \bibinfo {author} {\bibfnamefont {A.~W.}\ \bibnamefont
  {Sandvik}},\ }\href {\doibase 10.1103/physreve.66.046701} {\bibfield
  {journal} {\bibinfo  {journal} {Physical Review E}\ }\textbf {\bibinfo
  {volume} {66}} (\bibinfo {year} {2002}),\
  10.1103/physreve.66.046701}\BibitemShut {NoStop}%
\bibitem [{\citenamefont {Ceperley}(1995)}]{Ceperley1995}%
  \BibitemOpen
  \bibfield  {author} {\bibinfo {author} {\bibfnamefont {D.~M.}\ \bibnamefont
  {Ceperley}},\ }\href {\doibase 10.1103/revmodphys.67.279} {\bibfield
  {journal} {\bibinfo  {journal} {Reviews of Modern Physics}\ }\textbf
  {\bibinfo {volume} {67}},\ \bibinfo {pages} {279} (\bibinfo {year}
  {1995})}\BibitemShut {NoStop}%
\bibitem [{\citenamefont {Pollock}\ and\ \citenamefont
  {Ceperley}(1987)}]{Pollock1987}%
  \BibitemOpen
  \bibfield  {author} {\bibinfo {author} {\bibfnamefont {E.~L.}\ \bibnamefont
  {Pollock}}\ and\ \bibinfo {author} {\bibfnamefont {D.~M.}\ \bibnamefont
  {Ceperley}},\ }\href {\doibase 10.1103/physrevb.36.8343} {\bibfield
  {journal} {\bibinfo  {journal} {Physical Review B}\ }\textbf {\bibinfo
  {volume} {36}},\ \bibinfo {pages} {8343} (\bibinfo {year}
  {1987})}\BibitemShut {NoStop}%
\bibitem [{\citenamefont {Gurarie}\ \emph {et~al.}(2009)\citenamefont
  {Gurarie}, \citenamefont {Pollet}, \citenamefont {Prokof'ev}, \citenamefont
  {Svistunov},\ and\ \citenamefont {Troyer}}]{Gurarie2009}%
  \BibitemOpen
  \bibfield  {author} {\bibinfo {author} {\bibfnamefont {V.}~\bibnamefont
  {Gurarie}}, \bibinfo {author} {\bibfnamefont {L.}~\bibnamefont {Pollet}},
  \bibinfo {author} {\bibfnamefont {N.~V.}\ \bibnamefont {Prokof'ev}}, \bibinfo
  {author} {\bibfnamefont {B.~V.}\ \bibnamefont {Svistunov}}, \ and\ \bibinfo
  {author} {\bibfnamefont {M.}~\bibnamefont {Troyer}},\ }\href {\doibase
  10.1103/physrevb.80.214519} {\bibfield  {journal} {\bibinfo  {journal}
  {Physical Review B}\ }\textbf {\bibinfo {volume} {80}} (\bibinfo {year}
  {2009}),\ 10.1103/physrevb.80.214519}\BibitemShut {NoStop}%
\bibitem [{\citenamefont {Vojta}\ and\ \citenamefont
  {Hoyos}(2014)}]{Vojta2014}%
  \BibitemOpen
  \bibfield  {author} {\bibinfo {author} {\bibfnamefont {T.}~\bibnamefont
  {Vojta}}\ and\ \bibinfo {author} {\bibfnamefont {J.~A.}\ \bibnamefont
  {Hoyos}},\ }\href {\doibase 10.1103/physrevlett.112.075702} {\bibfield
  {journal} {\bibinfo  {journal} {Physical Review Letters}\ }\textbf {\bibinfo
  {volume} {112}} (\bibinfo {year} {2014}),\
  10.1103/physrevlett.112.075702}\BibitemShut {NoStop}%
\bibitem [{\citenamefont {Vojta}\ \emph {et~al.}(2016)\citenamefont {Vojta},
  \citenamefont {Crewse}, \citenamefont {Puschmann}, \citenamefont {Arovas},\
  and\ \citenamefont {Kiselev}}]{Vojta2016}%
  \BibitemOpen
  \bibfield  {author} {\bibinfo {author} {\bibfnamefont {T.}~\bibnamefont
  {Vojta}}, \bibinfo {author} {\bibfnamefont {J.}~\bibnamefont {Crewse}},
  \bibinfo {author} {\bibfnamefont {M.}~\bibnamefont {Puschmann}}, \bibinfo
  {author} {\bibfnamefont {D.}~\bibnamefont {Arovas}}, \ and\ \bibinfo {author}
  {\bibfnamefont {Y.}~\bibnamefont {Kiselev}},\ }\href {\doibase
  10.1103/physrevb.94.134501} {\bibfield  {journal} {\bibinfo  {journal}
  {Physical Review B}\ }\textbf {\bibinfo {volume} {94}} (\bibinfo {year}
  {2016}),\ 10.1103/physrevb.94.134501}\BibitemShut {NoStop}%
\bibitem [{\citenamefont {Lopes}\ \emph {et~al.}(2017)\citenamefont {Lopes},
  \citenamefont {Eigen}, \citenamefont {Navon}, \citenamefont {Cl{\'{e}}ment},
  \citenamefont {Smith},\ and\ \citenamefont {Hadzibabic}}]{Lopes2017}%
  \BibitemOpen
  \bibfield  {author} {\bibinfo {author} {\bibfnamefont {R.}~\bibnamefont
  {Lopes}}, \bibinfo {author} {\bibfnamefont {C.}~\bibnamefont {Eigen}},
  \bibinfo {author} {\bibfnamefont {N.}~\bibnamefont {Navon}}, \bibinfo
  {author} {\bibfnamefont {D.}~\bibnamefont {Cl{\'{e}}ment}}, \bibinfo {author}
  {\bibfnamefont {R.~P.}\ \bibnamefont {Smith}}, \ and\ \bibinfo {author}
  {\bibfnamefont {Z.}~\bibnamefont {Hadzibabic}},\ }\href {\doibase
  10.1103/physrevlett.119.190404} {\bibfield  {journal} {\bibinfo  {journal}
  {Physical Review Letters}\ }\textbf {\bibinfo {volume} {119}} (\bibinfo
  {year} {2017}),\ 10.1103/physrevlett.119.190404}\BibitemShut {NoStop}%
\bibitem [{\citenamefont {Mukherjee}\ \emph {et~al.}(2017)\citenamefont
  {Mukherjee}, \citenamefont {Yan}, \citenamefont {Patel}, \citenamefont
  {Hadzibabic}, \citenamefont {Yefsah}, \citenamefont {Struck},\ and\
  \citenamefont {Zwierlein}}]{Mukherjee2017}%
  \BibitemOpen
  \bibfield  {author} {\bibinfo {author} {\bibfnamefont {B.}~\bibnamefont
  {Mukherjee}}, \bibinfo {author} {\bibfnamefont {Z.}~\bibnamefont {Yan}},
  \bibinfo {author} {\bibfnamefont {P.~B.}\ \bibnamefont {Patel}}, \bibinfo
  {author} {\bibfnamefont {Z.}~\bibnamefont {Hadzibabic}}, \bibinfo {author}
  {\bibfnamefont {T.}~\bibnamefont {Yefsah}}, \bibinfo {author} {\bibfnamefont
  {J.}~\bibnamefont {Struck}}, \ and\ \bibinfo {author} {\bibfnamefont {M.~W.}\
  \bibnamefont {Zwierlein}},\ }\href {\doibase 10.1103/physrevlett.118.123401}
  {\bibfield  {journal} {\bibinfo  {journal} {Physical Review Letters}\
  }\textbf {\bibinfo {volume} {118}} (\bibinfo {year} {2017}),\
  10.1103/physrevlett.118.123401}\BibitemShut {NoStop}%
\end{thebibliography}

%

\end{document}